\documentclass[12pt]{article}
\usepackage[left=2cm,top=2cm,right=2cm,bottom=2cm]{geometry}

\usepackage[utf8]{inputenc}
\usepackage[british]{babel}

\usepackage{amsfonts,amsmath,amssymb,graphicx,latexsym,xcolor,mathtools}
\usepackage[scr=boondox]{mathalpha}
\usepackage{hyperref,enumitem,enumerate,float}
\usepackage{amsthm,verbatim}
\usepackage{cite}
\usepackage[autostyle]{csquotes}

\numberwithin{equation}{section}
\pdfoutput=1

\theoremstyle{plain}

\theoremstyle{definition}

\newtheorem{defi/}[theorem]{Definition}
\newtheorem{rema/}[theorem]{Remark}
\newtheorem{exa/}[theorem]{Example}

\makeatletter
\newcommand*{\transpose}{%
	{\mathpalette\@transpose{}}%
}
\newcommand*{\@transpose}[2]{%
	\raisebox{\depth}{$\m@th#1\intercal$}%
}
\makeatother

\newcommand{\tensor}[1]{\underline{\underline{\boldsymbol{#1}}}}

\newcommand{\vectstr}[1]{\textbf{#1}}
\newcommand{\vect}[1]{\boldsymbol{#1}}
\newcommand{\tensorind}[2]{\underline{\underline{\boldsymbol{#1}_{#2}}}}
\newcommand{\tensorsup}[3]{\underline{\underline{\boldsymbol{#1}_{#2}^{#3}}}}

\definecolor{myred}{RGB}{160,0,0}
\definecolor{mygreen}{RGB}{0,160,0}
\definecolor{myblue}{RGB}{0,0,160}
\hypersetup{
	colorlinks = true,
	linkcolor = {myred},
	citecolor = {mygreen},
	urlcolor  = {myblue},
}

\title{Effective mass density for wave propagation in periodic layered media in the elasto-acoustic transition}

\author{Gabriel N\'u\~nez, William J. Parnell and Rapha\"{e}l C. Assier\\
	\footnotesize{Department of Mathematics, University of Manchester, Oxford Road, Manchester, {\rm M13 9PL}, UK}
}

\date{}

\begin{document}

\maketitle

\begin{abstract}
This article investigates the effective mass density for acoustic and elastic wave propagation in layered media, with emphasis on its transition between these regimes. Conventionally, it is possible to recover the governing equations of acoustics from those of elasticity via the no-shear limit. However, when considering an effective medium, the anisotropic effective mass density typically obtained for acoustics differs from the isotropic one found in elasticity. Furthermore, direct investigation of this limit is hindered by the fact that the effective mass density is independent of the shear modulus $\mu$. 
To tackle this problem, a transfer matrix approach is combined with Bloch's theorem to derive the effective wavenumber for a periodic material whose unit cell consists of two layers with different densities. The effective mass density is obtained analytically from the wavenumber, allowing for its evaluation in both regimes. Moreover, for the first time, a description of the transition from isotropic elastic to anisotropic acoustic effective density in a periodic layered medium is provided. Additionally, the existence of band structure in $\mu$-space is demonstrated, along with exceptional points, where compressional and shear modes coalesce and waves become evanescent, which heavily affects the behaviour of the effective density in the elasto-acoustic transition.
\end{abstract}

\section{Introduction}

It is well-known that, for various physical phenomena such as acoustic and elastic wave propagation, several types of complex and inhomogeneous materials can be represented as equivalent media with effective physical properties \cite{Sanchez83, Auriault91}.
Within this framework, the concept of effective mass density constitutes an interesting source of discussion due to its discrepancy between elastic and acoustic waves.
Regarding the elastic case, a \textit{static} effective mass density can be obtained as the arithmetic average of the densities of its constituents, i.e.,  $\rho_{\textnormal{eff}} = \varphi\rho_1+(1-\varphi)\rho_2$ for a two-component (isotropic) mixture, where $\rho_1$ and $\rho_2$ are the densities of the first and second components, and $\varphi$ represents the volume fraction of the first.
However, in the case of acoustics, it was first noted by Ament \cite{Ament53} that, for low-frequency acoustic waves, the effective mass density does not correspond to the static mass density, and which is often called the \textit{dynamic} effective density.
Later works have reached different approximations where the effective density for acoustics does not equal the static one \cite{kuster1974,berryman1980}.
Currently, it is commonly assumed that for elastic mixtures the correct representation corresponds to the static effective density, whereas for acoustic composites it is the dynamic effective density \cite{Mei07, Alam19}.
In this context, layered materials are of significant interest.
They have been extensively studied \cite{Thomson50,Graf,Brekhovskikh,Whitham,Kinsler,Blackstock,Anselmet,Royer,Wei}, and it has been mathematically proven that they can be represented via an effective description \cite{Schoenberg83,Fish01,Andrianov08}.  
The particular interest in these materials arises from the fact that, for elastic waves, their effective density is isotropic \cite{Postma55}, while in acoustics, the effective density is anisotropic and corresponds to the arithmetic average for waves propagating perpendicular to the layer boundaries and to the geometric average parallel to them \cite{Schoenberg83}.

Conventionally, at the level of governing equations, the formulation for elasticity can be reduced to that of acoustics by taking the limit in which shear effects vanish, i.e., $\mu \rightarrow 0$ where $\mu$ is the shear modulus.
This presents the problem that, due to the discrepancy in the effective mass density between the acoustic and elastic regimes, and since in elasticity the classical formulae for this parameter do not explicitly depend on the shear modulus $\mu$, it is not possible to recover the acoustic effective density by simply taking the no-shear limit in the elastic model.
This problem has recently been addressed by Núñez et al. \cite{Nunez25}, who demonstrated the transition of the effective mass density between the elastic and acoustics regimes as $\mu \rightarrow 0$.
The authors employed a generalised self-consistent method \cite{yang2003} to numerically derive a dynamic effective density for a two-layered medium, also demonstrating that the elastic effective mass density becomes anisotropic for small values of the shear modulus.
However, this work presented such transition for a single unit cell, while its behaviour for a periodic medium remains an open question.
Therefore, the main objective of the present work is to characterise said transition in the periodic case.

The zero-shear limit problem in periodic layered media is addressed by employing a transfer matrix approach together with Bloch's theorem to derive a $\mu$-dependent effective mass density of a two-phase layered medium. 
The transfer matrix method \cite{Thomson50,Brekhovskikh,Folds77,Scharnhorst83,Brouard95,Mead96,Allard,Jimenez} corresponds to a mathematical technique commonly employed to describe wave propagation in stratified or layered media.
It models wave behaviour in each layer through a transfer matrix that relates the relevant physical quantities at one boundary of the layer to those at the opposite boundary. 
By combining the transfer matrices of individual layers, a global transfer matrix can be constructed to characterise wave propagation through the entire layered material.
Subsequently, the periodicity is introduced via Bloch's theorem, which allows to study wave propagation in periodic media \cite{Brillouin,Sigalas05,Wen05,Phani06,Hussein09,Srikantha11}, and particularly periodic elastic composites \cite{Kushwaha93,Kushwaha94}.
For this, the transfer matrix of a single unit cell, comprising any number of individual components, can be used to characterise wave propagation in a medium with infinitely repeating unit cells when combined with Bloch's theorem.
This process leads to an eigenvalue problem whose solution allows to obtain the Bloch wavenumber and dispersion curves of the periodic medium. 
This approach has been successfully employed to investigate wave propagation in different types of  periodic media such as periodic engineering structures \cite{Mead96}, waveguides \cite{Bradley94}, flexural beams \cite{Yu06,Liu12,Zhang18}, multi-periodic acoustic composites \cite{Lee10}, metamaterials \cite{Lee18}, and, more relevantly for the present work, layered or stratified media \cite{Esquivel94,Cao95,Hussein06,Lee15,Alizadeh25}.

This methodology leads to the identification of pass bands and band gaps, which correspond to frequency ranges where wave propagation in the periodic material is either allowed or forbidden, respectively \cite{Kushwaha93,Kushwaha94}.
Although band gaps are usually defined in frequency, in this work what will be referred to as $\mu$-gaps are identified, corresponding to intervals of values of $\mu$ for a fixed frequency where wave propagation in the medium is forbidden.
Additionally, due to coupling between compressional and shear waves, brought about by mode conversion at the boundaries between layers at oblique incidence, the appearance of points in $\mu$-space where there is a strong interaction between wave modes, particularly exceptional points (EPs) and avoided crossings (ACs), is expected  \cite{Kato,Heiss90,Heiss00,Dembowski01,Wu04,Seyranian05,Heiss12}.
The EPs correspond to values of $\mu$ where the eigenvalues and associated eigenvectors of the transfer matrix coalesce, and the matrix becomes non-diagonalisable.
The ACs correspond to points where the eigenvalue branches repel each other, also termed as level repulsion in literature \cite{Wu04}.
These points have been thoroughly investigated in the context of composite materials \cite{Wu04,Lustig19,Alizadeh25}, phononic crystals \cite{Christensen16,Lu18}, waveguides \cite{Xiong16,Xiong17,Nennig20,Lawrie22}, and metamaterials \cite{Xiong17,Achilleos17,Wang22}.
More recently, they have received a lot of attention in non-Hermitian and PT-symmetric systems \cite{Heiss12,Zhu14,Christensen16,Achilleos17,Alizadeh25}.
Directly relevant to this article are the works carried out by Lustig et al. \cite{Lustig19} and Alizadeh and Amirkhizi \cite{Alizadeh25}, in which the authors have identified the existence of exceptional points in lossless elastic layered materials.

From the Bloch wavenumber of the layered material, an effective phase speed can be directly obtained.
This phase speed is equivalent to the phase speed of a homogenised medium with an effective mass density that can be anisotropic.
Thus, from this effective phase speed it is possible to obtain expressions for the effective properties of the material, particularly the effective mass density.
For this, it is necessary to characterise wave propagation in a medium with anisotropic mass density.
In both acoustics and elasticity, such media have been extensively studied in the context of metamaterials, particularly pertaining to the concept of cloaking \cite{Schoenberg83,Torrent08,Cummer07,Norris08,Milton06,Milton07}, for which the anisotropic density is introduced directly in the governing equations in tensorial form. 
Particularly, recently developed models for the phase speed in acoustics and elasticity considering anisotropic mass density \cite{Jaberzadeh19,Sang19,Yang21,Nunez25} will be employed in this work.

This article is structured as follows. 
In Section 2, the transfer matrix for a two-layered medium for both acoustic and elastic waves considering oblique incidence is derived, including the calculation of the reflection and transmission coefficients for a given number of unit cells placed in between two identical half-spaces exhibiting anisotropic mass density.
Section 3 establishes Bloch's theorem for the problem, and demonstrates the calculation of the Bloch wavenumber from the eigenvalues of the transfer matrix for both types of waves.
Here, the existence of band gaps for shear waves as functions of the shear modulus is displayed. 
Moreover, the appearance of EPs and ACs in both the $\omega$- and $\mu$-spaces, where $\omega$ is the angular frequency, is explored. 
Lastly, Section 4 presents the derivation of the effective mass density for both acoustics and elasticity from the Bloch wavenumber.
Moreover, the transition between the two regimes as the shear modulus tends to zero is shown, identifying that the effective mass density exhibits discontinuities around $\mu$-gaps where compressional and shear modes interact.
Nonetheless, it is demonstrated that the elastic effective mass density does tend to its acoustic counterpart, albeit with the care of considering that $\mu$-gaps will continue to appear as $\mu \rightarrow 0$.
Furthermore, it is demonstrated that this approach leads to equivalent results as the dynamic self-consistent method employed in \cite{Nunez25} when considering a sufficiently large number of unit cells.

\section{Transfer matrix method for wave propagation in layered media} \label{Sec_1}

To study wave propagation in a periodic two-layered medium, it is first necessary to characterise its unit cell via its transfer matrix. 
Accordingly, this section presents the derivation of the transfer matrix for a single layer for both acoustic and elastic wave propagation, before demonstrating how the transfer matrices of the individual layers can be combined to obtain the total transfer matrix of the two-layered unit cell.
In addition, the reflection and transmission coefficients of such a layered medium embedded between two identical anisotropic half-spaces are derived.
These coefficients are required in the dynamic self-consistent method employed in Section \ref{Sec_III} to numerically determine the effective mass density. 
The surrounding half-spaces are assumed to be anisotropic, as the self-consistent method requires an anisotropic formulation to recover an anisotropic effective mass density \cite{Nunez25}.
%

\subsection{Transfer matrix method for acoustics} \label{Sec_II_tm_acoustics}

\begin{figure}[h!]
    \centering
    \includegraphics[width=0.65\linewidth]{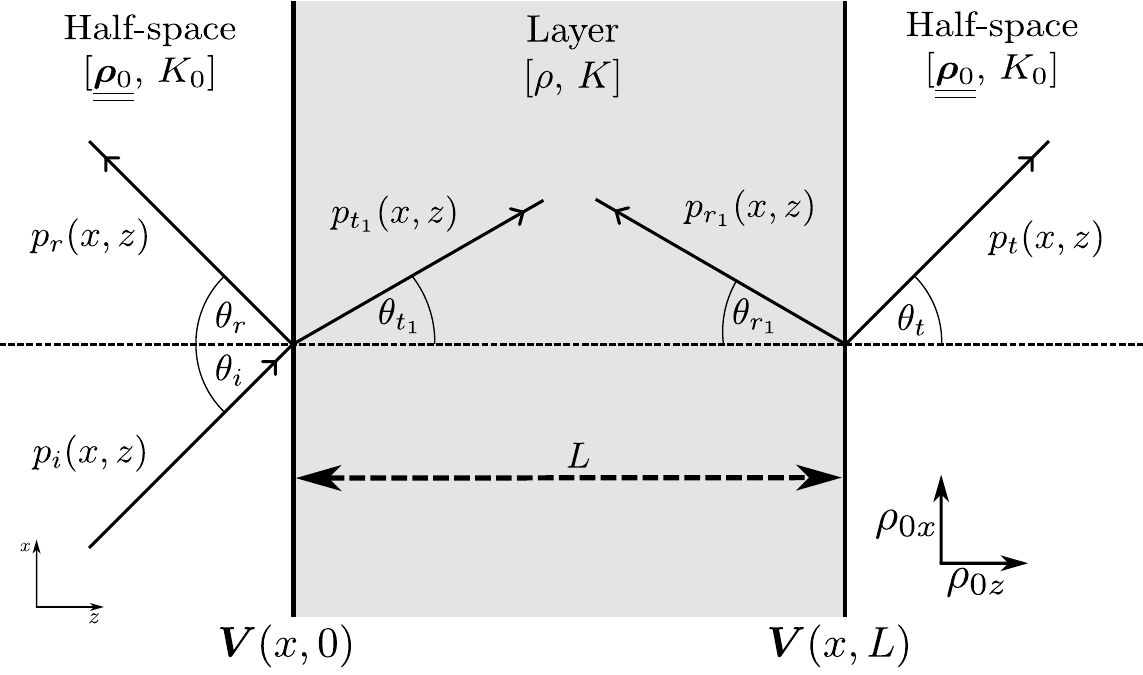}
    \caption{Graphical representation of acoustic wave propagation in a layer of material placed between two half-spaces with anisotropic mass density.}
    \label{Fig_acous_tm_1_layer}
\end{figure}

This section starts with the derivation of the transfer matrix for acoustic wave propagation in a layer of thickness $L$.
The derivation closely follows the approach presented in \cite{Allard,Jimenez}.
The layer is considered to be isotropic, characterised by an isotropic mass density $\rho$ and bulk modulus $K$.
The half-spaces have scalar bulk modulus $K_0$ and are assumed to exhibit anisotropic mass density $\tensorind{\rho}{0}$ as
\begin{equation} \label{Eq_density_tensor}
    \tensorind{\rho}{0} = \begin{pmatrix}
    \rho_{0x} & 0\\
    0 & \rho_{0z}
    \end{pmatrix},
\end{equation}
where $\rho_{0x}$ and $\rho_{0z}$ are the densities in the $x$- and $z$-directions, respectively.
A schematic of the geometry is shown in Figure \ref{Fig_acous_tm_1_layer}.

Employing the $\exp{(-i\omega t)}$ convention, the governing equations for time-harmonic acoustic wave propagation with angular frequency $\omega$ in the isotropic layer are the acoustic wave and Euler's equations, i.e.,
\begin{equation} \label{Eq_acous_wave_equation_aniso_density}
    \nabla \cdot \left( \frac{1}{\rho} \nabla p \right) + \frac{\omega^2}{K} p = 0,
    \quad 
    \vectstr{v} = - \frac{1}{i \omega \rho} \nabla p,
\end{equation}
where $p$ is the acoustic pressure and $\vectstr{v} = (v_x, v_z)$ is the acoustic velocity.
These are the physical quantities that will be related across the layer via the transfer matrix.

In the layer, the pressure field is expressed as $p (x,z) = \Bar{p}_{t_1} \exp{\left[i \mathcal{Z}^+(k,\theta_{t_1})\right]} + \Bar{p}_{r_1} \exp{\left[i \mathcal{Z}^-(k,\theta_{r_1})\right]}$, where $\mathcal{Z}^+(k,\theta) = k(x \sin{\theta} + z \cos{\theta})$ and $\mathcal{Z}^-(k,\theta) = k(x \sin{\theta} - z \cos{\theta})$, $\Bar{p}_{t_1}$ and $\Bar{p}_{r_1}$ are the wave amplitudes, and $k = \omega/c$ is the wavenumber, with phase speed $c = \sqrt{K/\rho}$.
This corresponds to a combination of two time-harmonic plane waves propagating in the $xz$-plane, one in the positive and the other in the negative $z$-direction.
Substituting this ansatz into the left equation in \eqref{Eq_acous_wave_equation_aniso_density} leads to an expression for the acoustic velocity.
In the following derivation, only the $z$-component of $\vectstr{v}$ is considered since the boundary conditions imposed at the interfaces enforce continuity of pressure and normal velocity.
Additionally, Snell's law implies that $\theta_{t_1} = \theta_{r_1}$, and so both angular dependencies are replaced by $\theta_l$ for clarity of notation.

The half-spaces in Figure \ref{Fig_acous_tm_1_layer} have no impact on the derivation of the transfer matrix, and thus are ignored for the time being.
The expressions for pressure and normal velocity in the layer are now rewritten in matrix form as
\begin{equation} \label{Eq_matrices_acous}
    \vect{V}(x,z) = \tensor{X} \ \tensor{D}(z) \ \vect{A} \exp{(i k x \sin{\theta_l})},
\end{equation}
where $\tensor{X}$ is the system matrix, $\tensor{D}(z)$ is a matrix of exponentials in $z$, $\vect{V}$ is the state vector and $\vect{A}$ is a vector containing wave amplitudes, i.e.,
\begin{align} \label{Eq_acous_state_amp}
\begin{split}
    &\tensor{X} = \begin{pmatrix}
    1 & 1 \\
    - 1/r & 1/r
    \end{pmatrix}, \ \tensor{D}(z) = \begin{pmatrix}
    \exp{\left[i k z \cos{\theta_l}\right]} & 0 \\
    0 & \exp{\left[-i k z \cos{\theta_l}\right]}
    \end{pmatrix}, \\
    &\vect{V}(x,z) =\begin{pmatrix}
    p(x,z)  \\
    v_z(x,z)
    \end{pmatrix}, \
    \vect{A} = \begin{pmatrix}
    \Bar{p}_{t_1}  \\
    \Bar{p}_{r_1}
    \end{pmatrix},
\end{split}
\end{align}
where $r = Z/ \cos{\theta_l}$ and $Z = \rho c$ is the acoustic characteristic impedance.
Since both $\tensor{X}$ and $\tensor{D}(z)$ are invertible, the pressure and normal velocity at $z = 0$ and $z = L$ are related as
\begin{equation} \label{Eq_acous_tm_layer_rel}
    \vect{V}(x,0) = \tensor{X} \ \tensor{D}^{-1}(L) \tensor{X}^{-1} \vect{V}(x,L),
\end{equation}
from which the transfer matrix is defined as 
\begin{equation} \label{Eq_acous_transf_mat_sin}
    \tensor{T} = \tensor{X} \ \tensor{D}^{-1}(L) \tensor{X}^{-1}.
\end{equation}
This expression yields the classical form of the transfer matrix for acoustic wave propagation through a layer as
\begin{equation} \label{Eq_acoustics_tm_1_layer}
    \tensor{T} = \begin{pmatrix}
    \cos{(k L \cos{\theta_l})} & i r \sin{(k L \cos{\theta_l})} \\
    i \dfrac{\sin{(k L \cos{\theta_l})}}{r} & \cos{(k L \cos{\theta_l})}.
    \end{pmatrix} \cdot
\end{equation}
The reflection and transmission coefficients $\mathscr{R}$ and $\mathscr{T}$ of the layer are now derived from $\tensor{T}$. 
To do so, the same geometry shown in Figure \ref{Fig_acous_tm_1_layer} is considered but focusing on the half-spaces to each side of the layer.
The governing equation for acoustic wave propagation in the half-spaces is found by replacing $1/\rho$ in \eqref{Eq_acous_wave_equation_aniso_density} by $\tensorind{\rho}{0}^{-1}$. 
As illustrated in the figure, a wave incident on the layer with angle $\theta_i$ from the left half-space with unit amplitude is considered.
Due to Snell's law, the propagation angles in the half-spaces are equal, i.e., $\theta_i = \theta_r = \theta_t$.
The pressure and normal velocity in the regions $z < 0$ and $z > L$ are written in terms of $\mathscr{R}$ and $\mathscr{T}$ as
\begin{equation} \label{Eq_acous_field_hsp}
    \vect{V}_0(x,z<0) = \tensorind{X}{0} \ \tensorind{D}{0}(z<0) \ \begin{pmatrix}
    1  \\
    \mathscr{R}
    \end{pmatrix} E(x), \quad
    \vect{V}_0(x,z>L) = \tensorind{X}{0} \ \tensorind{D}{0}(z>L) \ \begin{pmatrix}
    \mathscr{T}  \\
    0
    \end{pmatrix} E(x) ,
\end{equation}
where $E(x) = \exp{(i k_0(\theta_i) x \sin{\theta_i}}) $.
The matrices $\tensorind{X}{0}$ and $\tensorind{D}{0}(z)$ are given by \eqref{Eq_acous_state_amp} while replacing $k$ with $k_0 (\theta_i) = \omega/c_0 (\theta_i)$ and $r$ with $r_0(\theta_i) = Z_0(\theta_i)/\cos{\theta_i}$, and the angular-dependent acoustic characteristic impedance of the half-spaces is given by $Z_0(\theta_i) = \rho_{0z} c_0(\theta_i)$.
The angle-dependent phase speed is calculated as
\begin{equation} \label{Eq_acous_aniso_speed}
    c_0(\theta) = \sqrt{K_0\left(\frac{\sin^2{\theta}}{\rho_{0x}} + \frac{\cos^2{\theta}}{\rho_{0z}} \right)}.
\end{equation}

At the boundaries between the half-spaces and the layer, i.e., $z = 0$ and $z = L$, continuity of pressure and normal velocity is imposed, leading to
\begin{equation} \label{Eq_acous_tm_rt_cond}
    \vect{V}_0(x,0) = \vect{V}(x,0), \quad 
    \vect{V}_0(x,L) = \vect{V}(x,L),
\end{equation}
where $\vect{V}_0(x,z)$ is the state vector for the half-spaces.
Combining these boundary conditions with \eqref{Eq_acous_tm_layer_rel} and \eqref{Eq_acous_field_hsp} leads to a system of equations for the reflection and transmission coefficients as
\begin{equation} \label{Eq_acoustics_tm_ref_trans_sys}
    \begin{pmatrix}
    1  \\
    \mathscr{R}
    \end{pmatrix} = \tensor{M} \begin{pmatrix}
    \mathscr{T}  \\
    0
    \end{pmatrix}, \ \textnormal{where} \ \tensor{M} = \tensorind{X}{0}^{-1} \tensor{H} \ \textnormal{and} \ \tensor{H} = \tensor{T} \ \tensorind{X}{0} \ \tensorind{D}{0} (L).
\end{equation}
Solving for $\mathscr{T}$ and $\mathscr{R}$, the transmission and reflection coefficients of the layer are obtained as
\begin{equation} \label{Eq_acous_tm_R_T}
    \mathscr{T} = \frac{2 \exp{(-i k_0 h \cos{\theta_i})}}{T_{11} - T_{12}/r_0 - r_0 T_{21} + T_{22}}, \quad \mathscr{R} = \frac{T_{11} - T_{12}/r_0 + r_0 T_{21} - T_{22}}{T_{11} - T_{12}/r_0 - r_0 T_{21} + T_{22}}.
\end{equation}
where $T_{kl}$ are the components of $\tensor{T}$.

\begin{figure}[h!]
    \centering
    \includegraphics[width=0.55\linewidth]{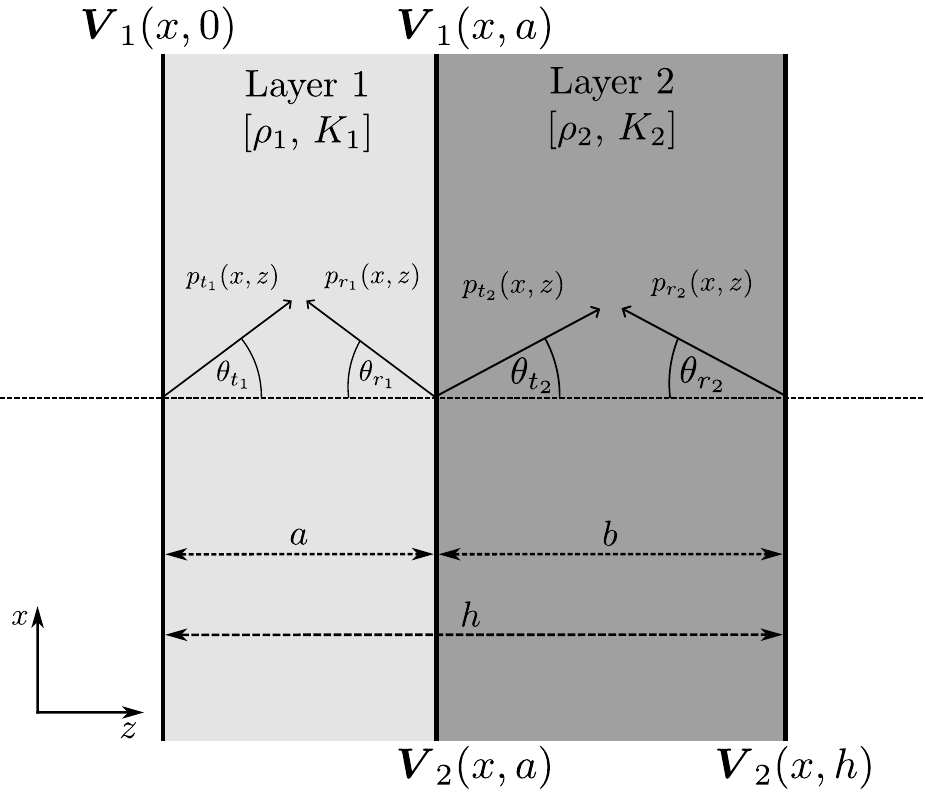}
    \caption{Graphical representation of acoustic wave propagation in two-layered material for transfer matrix method.}
    \label{Fig_acous_tm_2_layers}
\end{figure}

So far, the derivation has only led to the transfer matrix for a single homogeneous layer.
However, the focus of this section is on obtaining the transfer matrix for a medium composed of two distinct layers sharing a boundary that runs parallel to the $x$-axis, as illustrated in Figure \ref{Fig_acous_tm_2_layers}.
Layer 1 and Layer 2 are characterised by isotropic densities $\rho_1$ and $\rho_2$, and bulk moduli $K_1$ and $K_2$, respectively.
At the boundary between the two layers, continuity of pressure and normal velocity is imposed.
As in \eqref{Eq_acous_tm_rt_cond}, these boundary conditions are expressed as $\vect{V}_1(x,a) = \vect{V}_2(x,a)$, where $\vect{V}_1(x,a)$ and $\vect{V}_2(x,a)$ are the state vectors for the first and second layers, respectively, and $a$ is the thickness of the first layer.
From \eqref{Eq_acous_tm_layer_rel}, the pressure and normal velocity at the layer boundaries can be related via the transfer matrices $\tensorind{T}{1}$ and $\tensorind{T}{2}$ for each layer as $\vect{V}_1(x,0) = \tensorind{T}{1} \vect{V}_1(x,a)$ and $\vect{V}_2(x,a) = \tensorind{T}{2} \vect{V}_2(x,h)$,
with $h$ representing the sum of the thicknesses of the two layers.
From these equations, the total transfer matrix of the medium is obtained as $\tensorind{T}{t} = \tensorind{T}{1} \ \tensorind{T}{2}$.

From the transfer matrix of the medium, and treating the two-layered material as a single unit cell, it is straightforward to compute the transfer matrix for any given number of unit cells.
Following the same reasoning as for $\tensorind{T}{t}$, the total transfer matrix $\tensorsup{T}{t}{(n)}$ for $n$ unit cells is given by
\begin{equation} \label{Eq_acous_total_tm_n_uc}
    \tensorsup{T}{t}{(n)} = \left(\tensorind{T}{1} \ \tensorind{T}{2}\right)^n,
\end{equation}
where in general the product between the individual transfer matrices is not commutative. 
From this result along with \eqref{Eq_acous_tm_R_T}, the reflection and transmission coefficients for an arbitrary number of unit cells can be calculated.

\subsection{Transfer matrix method for elasticity} 
\label{Sec_II_tm_elasticity}

\begin{figure}[h!]
    \centering
    \includegraphics[width=0.65\linewidth]{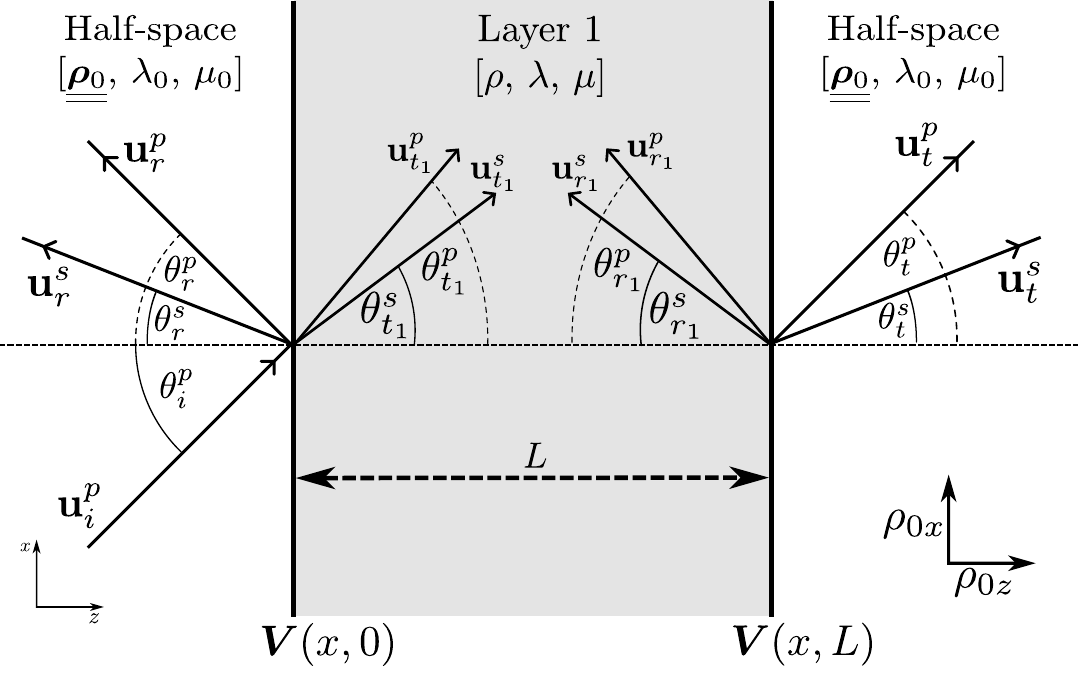}
    \caption{Graphical representation of elastic wave propagation in a layer of material placed between two half-spaces with anisotropic mass density.}
    \label{Fig_elas_tm_1_layer}
\end{figure}

This section closely follows the procedure presented in Section \ref{Sec_II_tm_acoustics}, but now focuses on elastic wave propagation in an isotropic layer of thickness $L$, density $\rho$, and Lam\'{e} parameters $\lambda$ and $\mu$.
A sketch of the set-up is shown in Figure \ref{Fig_elas_tm_1_layer}, where the main difference with the acoustic case is that now two types of waves can propagate in each direction, i.e., compressional (p) and shear (s) waves.
Moreover, when a wave impinges on a boundary at an angle, a phenomenon known as mode conversion takes place \cite{Anselmet,Graf}, in which an incident compressional elastic wave splits into compressional and shear waves.

The governing equation for elastic wave propagation in the layer is the time-harmonic Navier-Lam\'{e} equation, i.e.,
\begin{equation} \label{Eq_lin_elasticity_aniso_harmonic}
    \left( \lambda + \mu \right) \nabla \left( \nabla \cdot \vect{u} \right) + \mu \nabla^2 \vect{u} + \omega^2 \rho \vect{u} = \vect{0},
\end{equation}
where $\vect{u} = (u_x, u_z)$ corresponds to the displacement vector.
For the boundaries, perfect contact between layers is considered, represented by continuity of displacement and continuity of the traction vector $\vect{t} = \tensor{\sigma} \cdot \vect{n}$, which are the components of the second-rank stress tensor normal to the boundaries, i.e., $\sigma_{zz}$ and $\sigma_{xz}$.
The stress tensor $\tensor{\sigma}$ and second-rank infinitesimal strain tensor $\tensor{\varepsilon}\left(\vect{u}\right)$ for an elastic isotropic medium are defined as
\begin{equation} \label{Eq_elasticity_isotropic_stress_tensor}
    \tensor{\sigma} = \lambda \mathrm{Tr}(\tensor{\varepsilon}) \tensor{I} + 2 \mu \tensor{\varepsilon}, \quad
    \tensor{\varepsilon}(\vect{u}) = \frac{1}{2} \left( \nabla \vect{u} + (\nabla \vect{u})^\transpose \right),
\end{equation}
where $\tensor{I}$ is the second-order identity tensor and $^\transpose$ denotes transposition. 

The total displacement field in the layer is given by the sum of the displacements produced by compressional and shear waves propagating in the positive and negative $z$-directions, i.e., ${\vect{u} (x,z) = \vect{u}_{t_1}^p(x,z) + \vect{u}_{t_1}^s(x,z) + \vect{u}_{r_1}^p(x,z) + \vect{u}_{r_1}^s(x,z)}$, with
\begin{equation} \label{Eq_gen_disp_elas_trans}
    \vect{u}_{t_1}^p(x,z) = \Bar{u}_{t_1}^p \vect{P}_{p}^+ \exp{\left[i \mathcal{Z}^+(k_{p},\theta_t^p)\right]}, \quad \vect{u}_{t_1}^s(x,z) = \Bar{u}_{t_1}^s \vect{P}_{s}^+ \exp{\left[i \mathcal{Z}^+(k_{s},\theta_t^s)\right]},
\end{equation}
\begin{equation} \label{Eq_gen_disp_elas_reflec}
    \vect{u}_{r_1}^p(x,z) = \Bar{u}_{r_1}^p \vect{P}_{p}^- \exp{\left[i \mathcal{Z}^-(k_{p},\theta_r^p)\right]}, \quad \vect{u}_{r_1}^s(x,z) = \Bar{u}_{r_1}^s \vect{P}_{s}^- \exp{\left[i \mathcal{Z}^-(k_{s},\theta_r^s)\right]},
\end{equation}
where $\Bar{u}_{t_1}^p$, $\Bar{u}_{t_1}^s$, $\Bar{u}_{r_1}^p$, and $\Bar{u}_{r_1}^s$ are the wave amplitudes.
The wavenumbers are given by $k_p = \omega /c_p$ and $k_s = \omega /c_s$, with $c_p = \sqrt{(\lambda + 2\mu)/\rho}$ and $c_s = \sqrt{\mu/\rho}$ denoting the compressional and shear wave phase speeds, respectively.
Moreover, $\vect{P}_{p}^\pm = (P_{px},\pm P_{pz})$ and $\vect{P}_{s}^\pm = (\pm P_{sx},P_{sz})$ are the polarisation vectors for each type of wave.
The components of the stress tensor are obtained by replacing the displacement fields of \eqref{Eq_gen_disp_elas_trans} and \eqref{Eq_gen_disp_elas_reflec} into \eqref{Eq_elasticity_isotropic_stress_tensor}.
For the following derivation, Snell's law is employed to write the propagation angles as $\theta^p_{t_1} = \theta^p_{r_1} = \theta_{p_1}$ and $\theta^s_{t_1} = \theta^s_{r_1} = \theta_{s_1}$ for clarity of notation.

As in acoustics, to derive the elastic transfer matrix $\tensor{T}$, the displacement and stress fields are written in matrix form. 
For this purpose, the system matrix $\tensor{X}$, state vector $\vect{V}(x,z)$, and amplitude vector $\vect{A}$ are defined as 
\begin{align} \label{Eq_elas_matrix_x}
\begin{split}
    &\tensor{X} = \begin{pmatrix}
    P_{p x} & P_{p x} & P_{sx} & -P_{s x} \\
    P_{p z} & - P_{p z} & P_{s z} & P_{s z} \\
    i k_p \xi^p(\theta_{p_1}) & i k_p \xi^p(\theta_{p_1}) & i k_s \xi^s(\theta_{s_1}) & -i k_s \xi^s(\theta_{s_1}) \\
    i k_p \chi^p(\theta_{p_1}) & - i k_p \chi^p(\theta_{p_1}) & i k_s \chi^s(\theta_{s_1}) & i k_s \chi^s(\theta_{s_1})
    \end{pmatrix}, \\ 
    &\vect{V}(x,z) = \begin{pmatrix}
    u_x(x,z) \\
    u_z(x,z) \\
    \sigma_{zz}(x,z) \\
    \sigma_{xz}(x,z)
    \end{pmatrix}, \quad
    \vect{A}(x) = \begin{pmatrix}
   \Bar{u}_{t_1}^p \\
    \Bar{u}_{r_1}^p \\
    \Bar{u}_{t_1}^s \\
    \Bar{u}_{r_1}^s  
    \end{pmatrix},
\end{split}
\end{align}
where the functions $\xi^m(\theta)$ and $\chi^m(\theta)$ represent the contributions of each wave type to the stress in the medium and are given by
\begin{equation} \label{Eq_elas_xi_chi_m}
    \xi^m(\theta) = \lambda P_{m x} \sin{\theta} + (\lambda + 2\mu) P_{m z} \cos{\theta}, \quad
    \chi^m(\theta) = \mu (P_{m x} \cos{\theta} + P_{m z} \sin{\theta}),
\end{equation}
with $m = p, s$.
In addition, two diagonal matrices $\tensorind{D}{z}(z)$ and $\tensorind{D}{x}(x)$ that account for the exponential $z$- and $x$-dependence, respectively, are also written as
\begin{equation} \label{Eq_elas_matrix_d_z}
    \tensorind{D}{z}(z) =\mathrm{diag}\left(\exp{\left[i k_p z \cos{\theta_{p_1}}\right]},\exp{\left[-i k_p z \cos{\theta_{p_1}}\right]},\exp{\left[i k_s z \cos{\theta_{s_1}}\right]},\exp{\left[-i k_s z \cos{\theta_{s_1}}\right]}\right),
\end{equation}
\begin{equation} \label{Eq_elas_matrix_d_x}
    \tensorind{D}{x}(x) =\mathrm{diag}\left(\exp{\left[i k_p x \cos{\theta_{p_1}}\right]},\exp{\left[i k_p x \cos{\theta_{p_1}}\right]},\exp{\left[i k_s x \cos{\theta_{s_1}}\right]},\exp{\left[i k_s x \sin{\theta_{s_1}}\right]}\right).
\end{equation}
The total field can now be expressed in matrix form as
\begin{equation}
    \vect{V}(x,z) = \tensor{X} \ \tensorind{D}{z}(z) \ \tensorind{D}{x}(x) \ \vect{A}.
\end{equation}

As in acoustics, the displacement and normal stress fields are evaluated at both boundaries of the layer, i.e., $z =0 $ and $z = L$.
The state vectors at each boundary are related as in \eqref{Eq_acous_tm_layer_rel}, and thus, considering both $\tensor{X}$ and $\tensorind{D}{z}(z)$ are invertible, it is direct to see that the transfer matrix is calculated by \eqref{Eq_acous_transf_mat_sin} while replacing $\tensor{D}(L)$ by $\tensorind{D}{z}(L)$.
Since the components of the elastic transfer matrix $\tensor{T}$ involve coupling between the wavenumbers, polarisation vectors, the functions $\xi^m(\theta)$ and $\chi^m(\theta)$, and trigonometric functions, they become too large and thus the full matrix is not given here.
Instead, the detailed form of $\tensor{T}$ is included in Appendix \ref{Append_A}.
Having derived the transfer matrix for the elastic layer, the next step is to obtain its reflection and transmission coefficients.
As in acoustics, a geometry consisting of an elastic layer sandwiched between two identical half-spaces exhibiting anisotropic mass density $\tensorind{\rho}{0}$ is considered, as shown in Figure \ref{Fig_elas_tm_1_layer}.
The governing equation for this medium is given by \eqref{Eq_lin_elasticity_aniso_harmonic} while replacing the scalar mass density with $\tensorind{\rho}{0}$.
This leads to angle-dependent phase speeds as \cite{Nunez25}
\begin{equation} \label{Eq_elasticity_aniso_p_wave_speed}
    c_p(\theta) = \sqrt{\frac{\Gamma_{22}(\theta) \mathcal{R}_\rho + \Gamma_{11}(\theta)  + \sqrt{\left(\Gamma_{22}(\theta) \mathcal{R}_\rho - \Gamma_{11}(\theta) \right)^2 + 4 \Gamma_{12}^2(\theta) \mathcal{R}_\rho}}{2 \rho_x}},
\end{equation}
\begin{equation} \label{Eq_elasticity_aniso_s_wave_speed}
    c_s(\theta) = \sqrt{\frac{\Gamma_{22}(\theta) \mathcal{R}_\rho + \Gamma_{11}(\theta)  - \sqrt{\left(\Gamma_{22}(\theta) \mathcal{R}_\rho - \Gamma_{11}(\theta) \right)^2 + 4 \Gamma_{12}^2(\theta) \mathcal{R}_\rho}}{2 \rho_x}},
\end{equation}
where $\mathcal{R}_\rho = \rho_x/\rho_z$ and
\begin{equation} \label{Eq_elasticity_aniso_christoffel_matrix_comp_1}
    \Gamma_{11}(\theta) = (\lambda + 2\mu) \sin^2{\theta} + \mu \cos^2{\theta}, \quad \Gamma_{22}(\theta) = (\lambda + 2\mu) \cos^2{\theta} + \mu \sin^2{\theta},
\end{equation}
\begin{equation*}
    \Gamma_{12}(\theta) = (\lambda + \mu) \cos{\theta} \sin{\theta}.
\end{equation*}
To determine the reflection and transmission coefficients $\mathscr{R}_m$ and $\mathscr{T}_m$ for compressional and shear waves, an incident wave of unit amplitude  is considered, propagating from the left half-space with angle $\theta_i$ with respect to $z$.
In principle, the incident field could be a compressional wave, a shear wave, or a combination of both.
However, for the present case, only an incident compressional wave is considered.
This choice is motivated by the fact that, in Section \ref{Sec_III}, the acoustic limit is taken, whereby compressional waves reduce into purely acoustic ones while shear effects vanish.
Due to Snell's law, it is found that $\theta_i = \theta^p_t = \theta^p_r = \theta_p$ and $\theta_t^s = \theta_r^s = \theta_s$.
Moreover, ${k_p(\theta_p) \sin{\theta_p} =  k_s(\theta_s) \sin{\theta_s}}$, and so for the following representation the exponential functions in $x$ are written in terms of $k_p$ and $\theta_p$.
The displacements and stresses in the left-hand side ($z < 0$) and right-hand side ($z > L$) half-spaces can be written directly in matrix form as
\begin{align}
\begin{split}
    \vect{V}_0(x,z<0) &= \tensorind{X}{0} \ \tensorind{D}{z_0}(z) \ \begin{pmatrix}
    1   \\
    \mathscr{R}_p \\
    0   \\
    \mathscr{R}_s
    \end{pmatrix} \exp{[i k_p(\theta_p) x \sin{\theta_p}]}, \\ 
    \vect{V}_0(x,z>L) &= \tensorind{X}{0} \ \tensorind{D}{z_0}(z) \ \begin{pmatrix}
    \mathscr{T}_p \\
    0   \\
    \mathscr{T}_s \\
    0
    \end{pmatrix} \exp{[i k_p(\theta_p) x \sin{\theta_p}]},
\end{split}
\end{align}
where $\tensorind{X}{0}$ and $\tensorind{D}{z_0}$ are obtained from \eqref{Eq_elas_matrix_x} and \eqref{Eq_elas_matrix_d_z}, respectively, by replacing the layer parameters with those of the half-spaces.
The boundary conditions at the boundaries between layer and half-spaces, i.e., continuity of displacement and normal stress, are expressed in vector form as in \eqref{Eq_acous_tm_rt_cond} while replacing the acoustic state vector with the elastic one.
Following the same procedure as in \eqref{Eq_acous_tm_rt_cond}--\eqref{Eq_acoustics_tm_ref_trans_sys}, the coefficients are obtained by solving the system
\begin{equation}
    \begin{pmatrix}
    1   \\
    \mathscr{R}_p \\
    0   \\
    \mathscr{R}_s
    \end{pmatrix} = \tensor{M} \begin{pmatrix}
    \mathscr{T}_p \\
    0   \\
    \mathscr{T}_s \\
    0
    \end{pmatrix}, \quad \textnormal{where} \ \tensor{M} = \tensorind{X}{0}^{-1} \tensor{H} \ \textnormal{and} \ \tensor{H} = \tensor{T} \ \tensorind{X}{0} \ \tensorind{D}{0} (L).
\end{equation}
Solving for the coefficients leads to
\begin{equation}
    \mathscr{R}_p = \frac{M_{33} M_{21} - M_{23} M_{31}}{D_e}, \quad \mathscr{R}_s = \frac{M_{31} M_{43} - M_{33} M_{41}}{D_e}, \quad \mathscr{T}_p = \frac{M_{33}}{D_e}, \quad \mathscr{T}_s = -\frac{M_{31}}{D_e},
\end{equation}
where $D_e = M_{11} M_{33} - M_{13} M_{31}$ and $M_{kl}$ are the components of $\tensor{M}$.
Following the same reasoning as in the previous section, and replacing the boundary conditions for acoustics with those relevant to elasticity, the total transfer matrix of a material composed of $n$ unit cells, each consisting of two layers with distinct physical properties, is given by \eqref{Eq_acous_total_tm_n_uc}.

\section{Bloch's theorem for periodic layered media}

\subsection{Bloch's theorem and eigenvalue analysis of transfer matrix} \label{Sec_IIa}

%
The transfer matrix approach is now combined with Bloch's theorem to model wave propagation in an infinitely periodic layered medium whose unit cell corresponds to the two-layered material discussed in the previous section.
The general Bloch's theorem for wave propagation across a single unit cell of thickness $h$ yields
\begin{equation} \label{Eq_bloch_theo}
    \vect{V}(x,0) = \exp{(i \mathcal{K}h)} \vect{V}(x,h),
\end{equation}
where $\vect{V}(x,0)$ and $\vect{V}(x,h)$ are the state vectors given in \eqref{Eq_acous_state_amp} for acoustics and \eqref{Eq_elas_matrix_x} for elasticity, and $\mathcal{K}$ denotes the Bloch wavenumber.
The transfer matrix $\tensor{T}$ of the unit cell relates the state vectors as $\vect{V}(x,0) = \tensor{T} \ \vect{V}(x,h)$.
Combining these two equations leads to 
\begin{equation}
    \left( \tensor{T} - \exp{(i \mathcal{K}h)\tensor{I}} \right) \vect{V}(x,h) = \vect{0}.
\end{equation}
This system has a non-trivial solution for $\vect{V}(x,h)$ only if 
\begin{equation} \label{Eq_bloch_eig_exp}
    \det{\left(\tensor{T} - \gamma \tensor{I}\right)} = 0, \ \text{where} \ \gamma = \exp{(i \mathcal{K}h)}.
\end{equation}
Solving this equation yields the eigenvalues $\gamma$ of the transfer matrix $\tensor{T}$ for the layered medium.
For acoustics, it is straightforward to obtain the eigenvalues of the transfer matrix by solving its second-order characteristic equation.
However, for elasticity, this characteristic equation corresponds to a fourth-order polynomial, i.e., four independent eigenvalues, and thus directly solving for its roots becomes challenging.
Nonetheless, the problem is significantly simplified considering the eigenvalues come in two reciprocal pairs as $\gamma_{1}$ and $\gamma_{1}^\ast = 1/\gamma_{1}$, and $\gamma_{2}$ and $\gamma_{2}^\ast = 1/\gamma_{2}$.
Introducing the auxiliary variables $\eta_{1,2} = \left( \gamma_{1,2} + 1/\gamma_{1,2}\right)$, it can be shown that the eigenvalues are given by
\begin{equation} \label{Eq_tm_eigs_eta_gamma}
    \gamma_{1,2} = \frac{\eta_{1,2} + i \sqrt{4 - \eta_{1,2}^2}}{2}, \quad \gamma_{1,2}^* = \frac{\eta_{1,2} - i \sqrt{4 - \eta_{1,2}^2}}{2}, \quad \eta_{1,2} = \frac{\mathscr{c}_1 \pm \sqrt{\mathscr{c}_1^2 - 4(\mathscr{c}_2 - 2)}}{2}, 
\end{equation}
where the positive sign before the square root corresponds to $\eta_1$ and the negative to $\eta_2$. The constants $\mathscr{c}_1$ and $\mathscr{c}_2$ are given by
\begin{equation} \label{Eq_constants_c1_c2}
    \mathscr{c}_1 = \mathrm{Tr}(\tensor{T}), \quad \mathscr{c}_2 = \frac{1}{2} \left( \left(\mathrm{Tr}(\tensor{T})\right)^2 - \mathrm{Tr}(\tensor{T}^2) \right). 
\end{equation}
This solution gives two complex-conjugate pairs of eigenvalues, resulting in four different Bloch wavenumbers $\mathcal{K}_{1,2}$ and $-\mathcal{K}_{1,2}$.
The details of the derivation are given in Appendix \ref{Append_B}. 

Regarding the Bloch wavenumber, from \eqref{Eq_bloch_eig_exp} it follows that \\ ${\eta_{1,2} = \exp{(i \mathcal{K}_{1,2}h)} + \exp{(-i \mathcal{K}_{1,2}h)} = 2 \cos{(\mathcal{K}_{1,2}h)}}$.
This representation directly leads to the identification of two regimes for each pair of eigenvalues.
For $|\eta_{1,2}| < 2$, the wavenumbers $\mathcal{K}_{1,2}$ become completely real and thus waves can propagate freely in the layered medium.
Here, it follows from \eqref{Eq_tm_eigs_eta_gamma} that, for $|\eta_{1,2}| < 2$, $|\gamma_{1,2}| = |\gamma_{1,2}^*| = 1$.
On the other hand, if $|\eta_{1,2}| > 2$, $|\gamma_{1,2}| \neq |\gamma_{1,2}^*| \neq 1$, leading to complex $\mathcal{K}_{1,2}$ and thus waves decay exponentially in the medium, corresponding to a band gap where wave propagation in the material is forbidden. 
Additionally, finding the roots of the function $f(\eta_{1,2}) = |\eta_{1,2}| - 2$ allows one to determine the edges of the band gaps, and it can be seen that these points are equivalent to a pair of real eigenvalues equal to $1$.
The previous analysis is valid for either pair of eigenvalues, and the auxiliary variables $\eta_{1,2}$ allow one to neatly separate them by grouping one of the eigenvalues with its respective reciprocal.
In that way, it is convenient to define a $\eta_p$ for compressional waves and $\eta_s$ for shear waves, which in turn leads to $\gamma_{p_1}$ and $\gamma_{p_2}$, and $\gamma_{s_1}$ and $\gamma_{s_2}$.
To accurately assign each eigenvalue pair with its respective wave mode, the eigenvectors of the transfer matrix are analysed.
These contain information about the displacement and normal stress on the boundary of the unit cell.
The components can be separated between those typically associated with longitudinal motion, namely $u_z$ and $\sigma_{zz}$, which come to represent the compressional waves propagating in the system, and those pertaining to transversal motion, i.e., $u_x$ and $\sigma_{xz}$ and correspond to shear waves.
Based on this, for the rest of this work the eigenvalues for which $|u_z| > |u_x|$ and $|\sigma_{zz}|>|\sigma_{xz}|$ are assigned as the compressional mode, while the opposite case corresponds to the shear mode.

An interesting case of study occurs when $\mathscr{c}_1^2 = 4(\mathscr{c}_2 - 2)$ in \eqref{Eq_tm_eigs_eta_gamma}.
In this case, the quadratic polynomial in \eqref{Eq_charac_polyn_eta} becomes single rooted, and $\eta_{1,2} = \mathscr{c}_1/2$.
These correspond to exceptional points or EPs for short, where eigenvalues representing distinct modes coalesce, meaning $\gamma_1 = \gamma_2$ and $\gamma_1^* = \gamma_2^*$.
This implies that the transfer matrix has only two complex-conjugate eigenvalues, each with algebraic multiplicity of 2.
It can be ensured that these points correspond to exceptional points because the eigenvectors coalesce, going from four linearly-independent eigenvectors to just two.
This results in the transfer matrix becoming non-diagonalisable at the EPs.
Furthermore, the modes become highly hybridised, meaning there is not a reliable approach to label the branches as either compressional or shear.
Additionally, if $4(\mathscr{c}_2 - 4) > \mathscr{c}_1^2$, $\eta_{1,2}$ becomes 
\begin{equation}
    \eta_{1,2} =  \frac{\mathscr{c}_1 \pm i \sqrt{4(\mathscr{c}_2 - 2) - \mathscr{c}_1^2}}{2},
\end{equation}
where $\eta_{1,2} \in \mathbb{C}$.
This in turn leads to complex eigenvalues with absolute values either smaller or larger than one, indicating the existence of coupled evanescent modes \cite{Alizadeh25}.
Moreover, EPs are expected to appear in pairs and demarcate an EP gap.
%
Additionally, the definition of EPs imposes no restriction on $|\eta_{1,2}|$ and thus these points can theoretically emerge within pass bands or band gaps. 
Furthermore, a band gap edge can potentially be inside of a EP gap, meaning that one of the EPs is located inside the band gap and the other outside.
This case will be explored in more detail in the next sections.
A particular case occurs when $\mathscr{c}_1^2 = 4(\mathscr{c}_2 - 2)$ and $|\eta_{1,2}| = 2$, which implies the merging of an EP with a band gap edge and the coalescing of all four eigenvalues of $\tensor{T}$ as $\gamma_{1,2} = \gamma_{1,2}^* = 1$.
Importantly, the existence of EPs is only ensured when the propagation angles in the layers are not zero.
When considering normal incidence, mode conversion is eliminated from the problem, meaning compressional and shear waves become fully decoupled. 
For the transfer matrix, the components in the first and last row become purely shear, while those in the second and third rows describe exclusively compressional waves.
This effectively decouples the $4 \times 4$ global elastic transfer matrix into two independent $2 \times 2$ ones.
In this case, the compressional and shear eigenvalues still exhibit crossing points.
However, for each crossing, the eigenvectors remain linearly independent, and thus they do not correspond to exceptional points.
These are known as Dirac or diabolical points \cite{Seyranian05}.

So far, the existence of EPs has only been studied for real-valued parameters. 
However, the equation $\mathscr{c}_1^2 = 4(\mathscr{c}_2 - 2)$ can also exhibit solutions in complexified parameter space.
These EPs happen in complex conjugate pairs and lead to the appearance of avoided crossings or ACs between the branches of the Riemann surfaces of $\eta_{1,2}$, and thus of $\gamma_{1,2}$ and $\gamma_{1,2}^*$.
In the ACs, the eigenvalues seem to be approaching a crossing point, only to ultimately repel each other.
Nonetheless, the absolute values of the components of the eigenvectors do exhibit a single crossing at ACs, as it will be demonstrated in the following examples.
This implies that, if for a particular eigenvalue the corresponding eigenvector satisfies that $|u_z| > |u_x|$ and $|\sigma_{zz}|>|\sigma_{xz}|$, after an avoided crossing the eigenvector will exhibit $|u_z| < |u_x|$ and $|\sigma_{zz}|<|\sigma_{xz}|$.
Therefore, labelling the eigenvalues based on the eigenvectors will lead to branch swapping when approaching ACs.
In this case, even though the eigenvalues themselves remain continuous, the physical interpretation of each branch switches, with the previously compressional branch becoming shear and vice versa. 
Importantly, the displacement and stress components of the eigenvectors do not necessarily cross at the same parameter values, as will be demonstrated in the next section. 

\subsection{Illustrative examples} \label{Sec_IIa2}

\begin{figure}[h!]
    \centering
    \includegraphics[width=\linewidth,trim={0 0.32cm 0 0},clip]{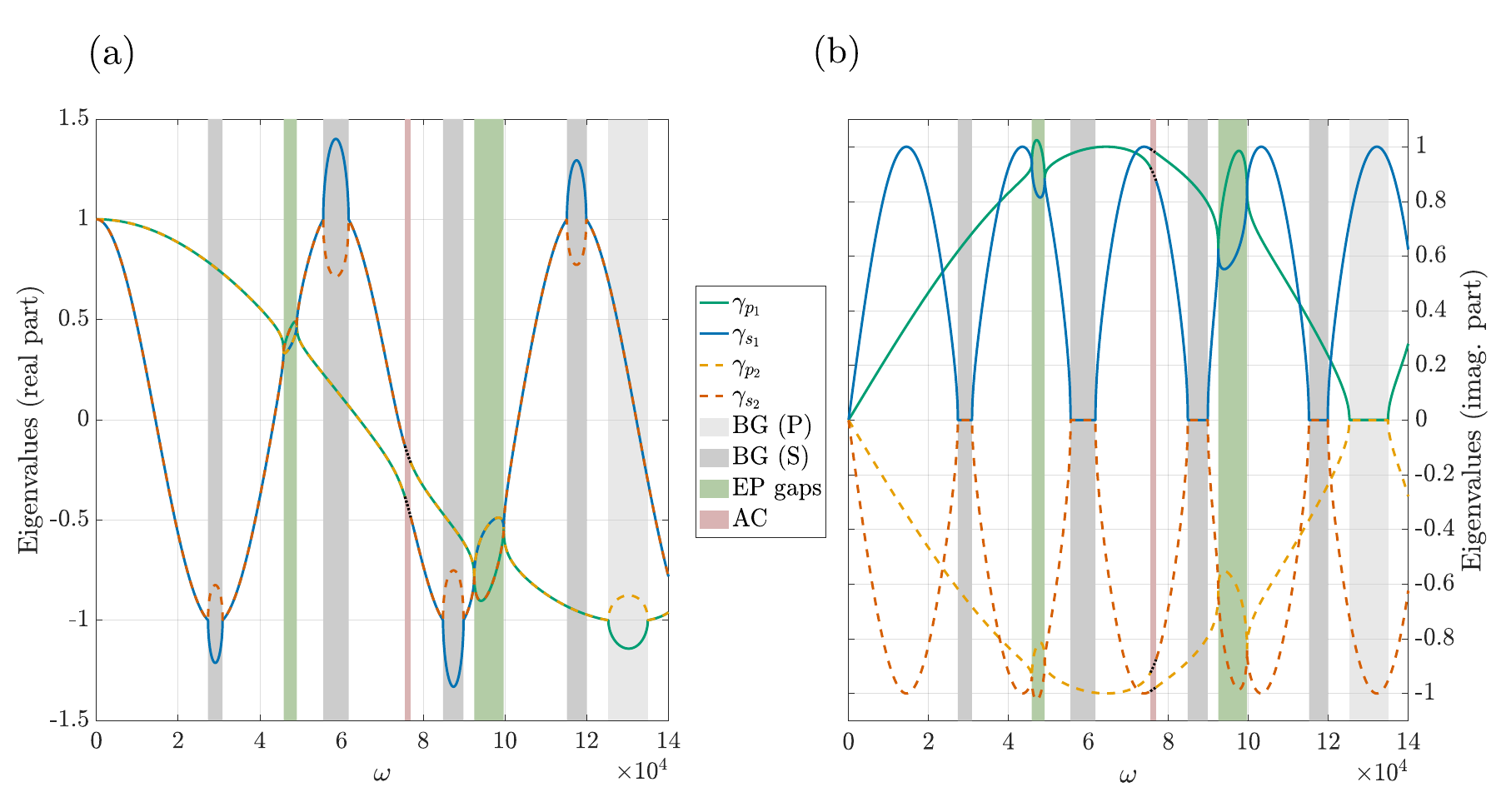}
    \caption{Real (a) and imaginary (b) parts of the eigenvalues associated with both compressional and shear waves against angular frequency.}
    \label{Fig_elas_tm_eigs_freq}
\end{figure}

\begin{figure}[h!]
    \centering
    \includegraphics[width=\linewidth,trim={0 0.32cm 0 0},clip]{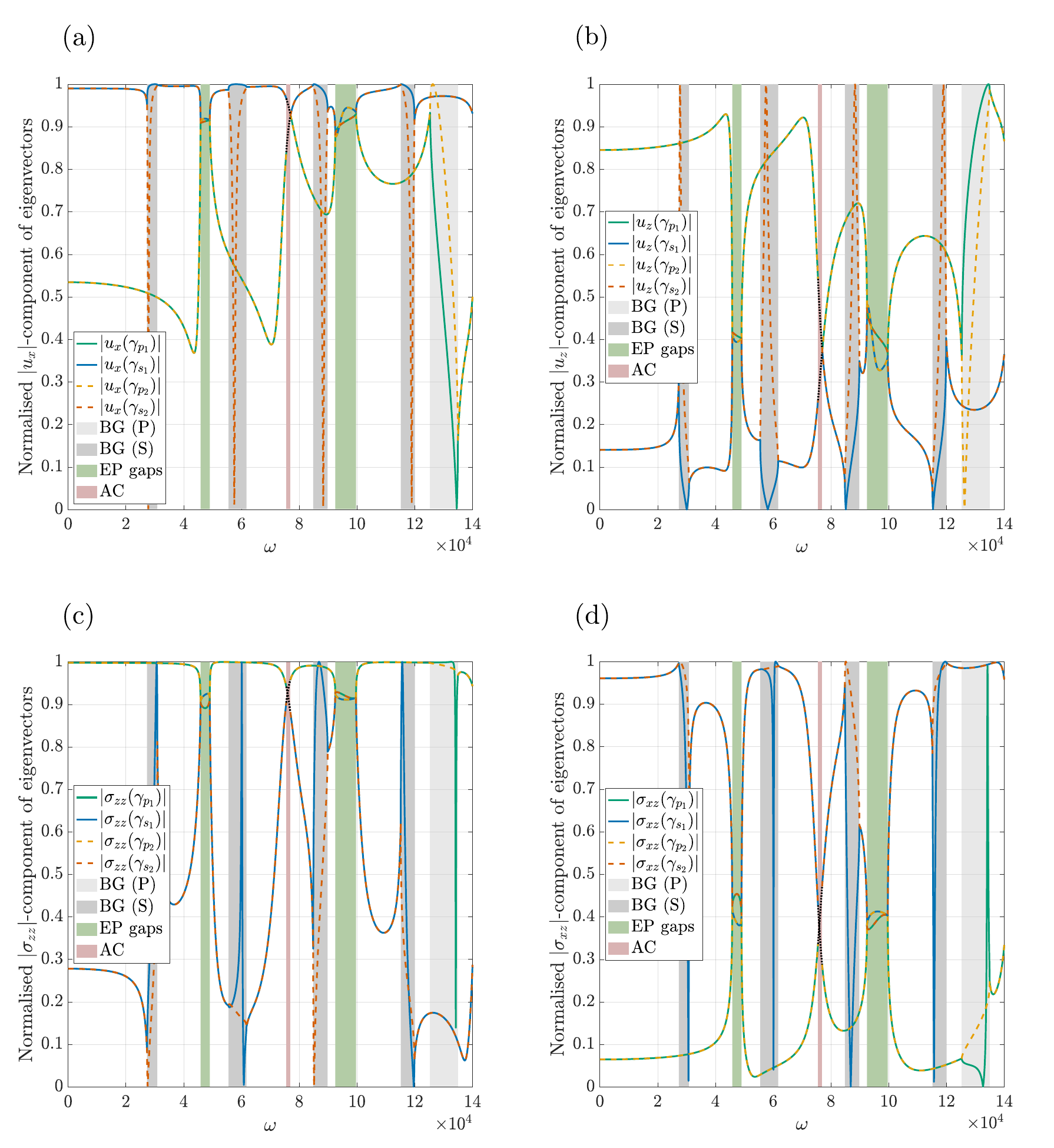}
    \caption{Absolute value of the displacement in the $x$- (a) and $z$-directions (b) and of normal stresses $\sigma_{zz}$ (c) and $\sigma_{xz}$ (d) associated with both compressional and shear waves against angular frequency.}
    \label{Fig_elas_tm_egvec_freq}
\end{figure}

\begin{figure}[h!]
    \centering
    \includegraphics[width=\linewidth,trim={0 0.32cm 0 0},clip]{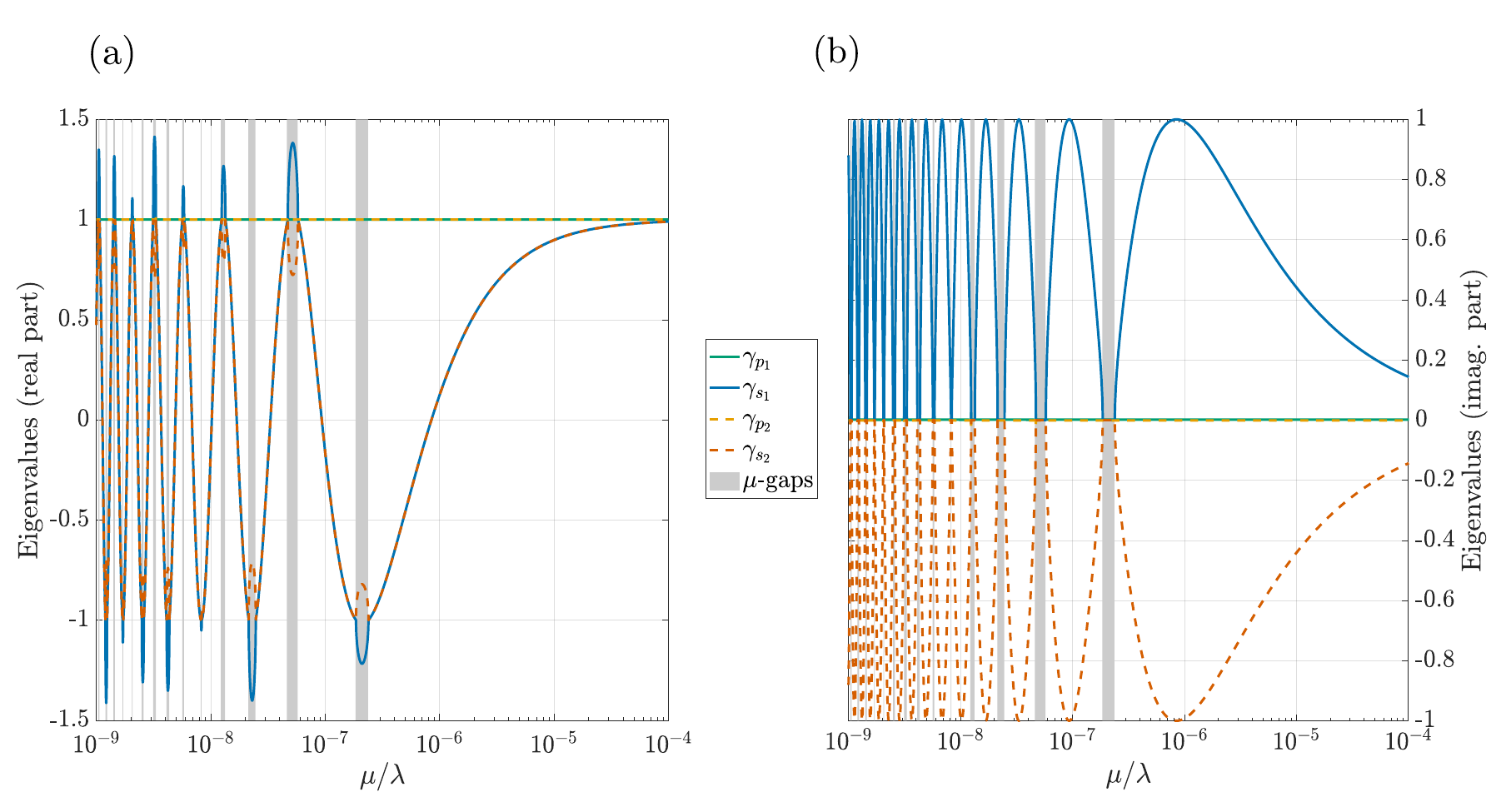}
    \caption{Real (a) and imaginary (b) parts of the eigenvalues associated with both compressional and shear waves against $\mu/\lambda$.}
    \label{Fig_elas_tm_eigs_mu}
\end{figure}

%
Even though it is of interest for this work to analyse the behaviour of the eigenvalues while the shear modulus tends to zero, it is illustrative to look at their behaviour as a function of frequency, mainly because this corresponds to the usual fashion in which dispersion curves are represented.
Moreover, the EP gaps and ACs are more conveniently spotted in this manner.
A plot of the eigenvalues for both compressional and shear modes against frequency for the two-layered unit cell of Figure \ref{Fig_acous_tm_2_layers}, is presented in Figure \ref{Fig_elas_tm_eigs_freq}, exhibiting their real (a) and imaginary (b) parts.
In this example, a propagation angle for compressional waves of $\theta_{p_1} = \pi/4$ in the first layer is considered, where the remaining angles are calculated via Snell's law as 
\begin{equation} \label{Eq_snells_law} 
    \frac{\sin{\theta_{p_1}}}{c_{p_1}}  = \frac{\sin{\theta_{p_2}}}{c_{p_2}} = \frac{\sin{\theta_{s_1}}}{c_{s_1}}  = \frac{\sin{\theta_{s_2}}}{c_{s_2}},
\end{equation}
where $c_{p_1}$ and $c_{p_2}$, and $c_{s_1}$ and $c_{s_2}$ are the compressional and shear phase speeds in layers 1 and 2, respectively, and $\theta_{p_2}$ and $\theta_{s_2}$ are the propagation angles for compressional and shear waves in the second layer.
The band gaps in frequency are marked with a dark grey box for shear waves and a light grey box for compressional waves, where $|\gamma_{s_{1,2}}| \neq 1$ and $|\gamma_{p_{1,2}}| \neq 1$, respectively.
These are identified as frequency intervals where the imaginary part of the eigenvalues is zero while their real part is either larger or smaller than unity.
Regarding the EPs, i.e., the edges of the green regions, the eigenvalues merge from two distinct pairs to just one conjugate pair. 
This is observed as a merging of all the real parts into a single one and a coalescing of the positive and negative imaginary parts.
Inside the EP gap, the eigenvalues exhibit an absolute value different than unity, as inside the band gaps.
In this case, the plot only shows EP gaps located in pass bands, since their behaviour inside a band gap is equivalent.
Lastly, highlighted in red are the avoided crossings.
Here the eigenvalues are approaching a crossing as frequency increases, to ultimately repel each other.
After the crossing, the physical nature of each branch, i.e., whether they correspond to compressional and shear modes, switches. 
These results are consistent with those presented previously in the literature \cite{Alizadeh25}.
To further elucidate the behaviour of elastic wave propagation in the layered medium, the eigenvectors of the transfer matrix against frequency are presented in Figure \ref{Fig_elas_tm_egvec_freq}.
For the displacement, each $\vect{u} (\gamma) = (u_x(\gamma), u_z(\gamma))$ is normalised by $\sqrt{|u_x(\gamma)|^2 + |u_z(\gamma)|^2}$.
In the case of the stress, each $\sigma_{zz}(\gamma)$ and $\sigma_{xz}(\gamma)$ are normalised by $\sqrt{|\sigma_{zz}(\gamma)|^2 + |\sigma_{xz}(\gamma)|^2}$.
In the plots, it can be identified that the branch labelled as compressional exhibits dominance of the $|u_z|$ and $|\sigma_{zz}|$ terms over most of the plotted frequency range.
Conversely, $|u_x|$ and $|\sigma_{xz}|$ are considerably larger for shear waves. 
However, at the EPs, the eigenvectors coalesce and the previously four independent eigenvectors become only two linearly independent ones, one for forward and the other one for backward propagating waves.
This is consistent with the analysis of the eigenvalues in Figure \ref{Fig_elas_tm_eigs_freq}.
On the other hand, studying the eigenvectors allows for further understanding of the behaviour of the mode switching at the avoided crossings.
As it can be observed in the figure, even though the eigenvalues repel each other, the eigenvectors do actually intersect at these avoided crossings. 
These intersections indicate that the previously dominantly compressional branch now exhibits a predominantly shear behaviour and vice versa.
This justifies the choice of discontinuously switching the labelling of the eigenvalues in Figure \ref{Fig_elas_tm_eigs_freq}.
Moreover, as noted previously, the components of the eigenvectors do not necessarily cross at the same frequency.
In particular, $|\sigma_{zz}|$ and $|\sigma_{xz}|$ intersect at a lower frequency, whereas $|u_x|$ and $|u_z|$ cross at a slightly higher frequency.
These plots also suggest that, around the EPs and ACs, the polarisation of the branches becomes highly mixed and thus the classification between compressional and shear modes becomes unreliable.
Next, a general example of the eigenvalues as $\mu \rightarrow 0$ is presented.
Figure \ref{Fig_elas_tm_eigs_mu} shows the real (a) and imaginary (b) parts of the eigenvalues associated with both compressional and shear waves, plotted as functions of the shear modulus $\mu$.
Importantly, for all the examples in $\mu$-space, a fixed frequency $\omega$ is considered.
Since in the next section the low-frequency effective mass density for wave propagation in this medium will be derived, this frequency is chosen so that the wavelength of the propagating waves is much larger than the thickness of the unit cell.
Regarding the compressional eigenvalues, it is observed that they remain nearly unchanged as $\mu \rightarrow 0$.
This is explained by the fact that, as $\mu$ decreases, the wavenumbers $k_{p_1} = \omega \sqrt{\rho_1/(\lambda+2\mu)}$ and ${k_{p_2} = \omega \sqrt{\rho_2/(\lambda+2\mu)}}$ of each layer approach constant values. 
Additionally, over the entire range of values of $\mu$ considered, the real parts of $\gamma_{p_1}$ and $\gamma_{p_2}$ remain close to unity, with the imaginary parts nearly vanishing.
In contrast, it can be seen that in the case of the shear eigenvalues $\gamma_{s_1}$ and $\gamma_{s_2}$, for values of $\mu$ close to $\lambda$, their behaviour resembles that of the compressional eigenvalues.
However, as $\mu$ decreases, the real parts begin to oscillate, with identical behaviour for both $\gamma_{s_1}$ and $\gamma_{s_2}$, while the imaginary part varies between $0$ and $1$ for $\gamma_{s_1}$ and between $-1$ and $0$ for $\gamma_{s_2}$.
Nevertheless, in certain ranges of $\mu$, the imaginary part vanishes and the eigenvalues leave the unit circle, meaning that their absolute value is different from one.
This behaviour coincides with that of the band gaps discussed in Figure \ref{Fig_elas_tm_eigs_freq}, only this time these are expressed in $\mu$-intervals instead of frequency ranges, i.e., $\mu$-gaps.
Additionally, following the behaviour observed in the frequency regime, it is expected that, around the $\mu$-gaps where $\gamma_s$ becomes close to 1, the compressional and shear branches will intersect, leading to the appearance of EPs and ACs.
This is investigated extensively in Section \ref{Sec_IIb}.

\begin{figure}[h!]
    \centering
    \includegraphics[width=\linewidth,trim={0 0.32cm 0 0},clip]{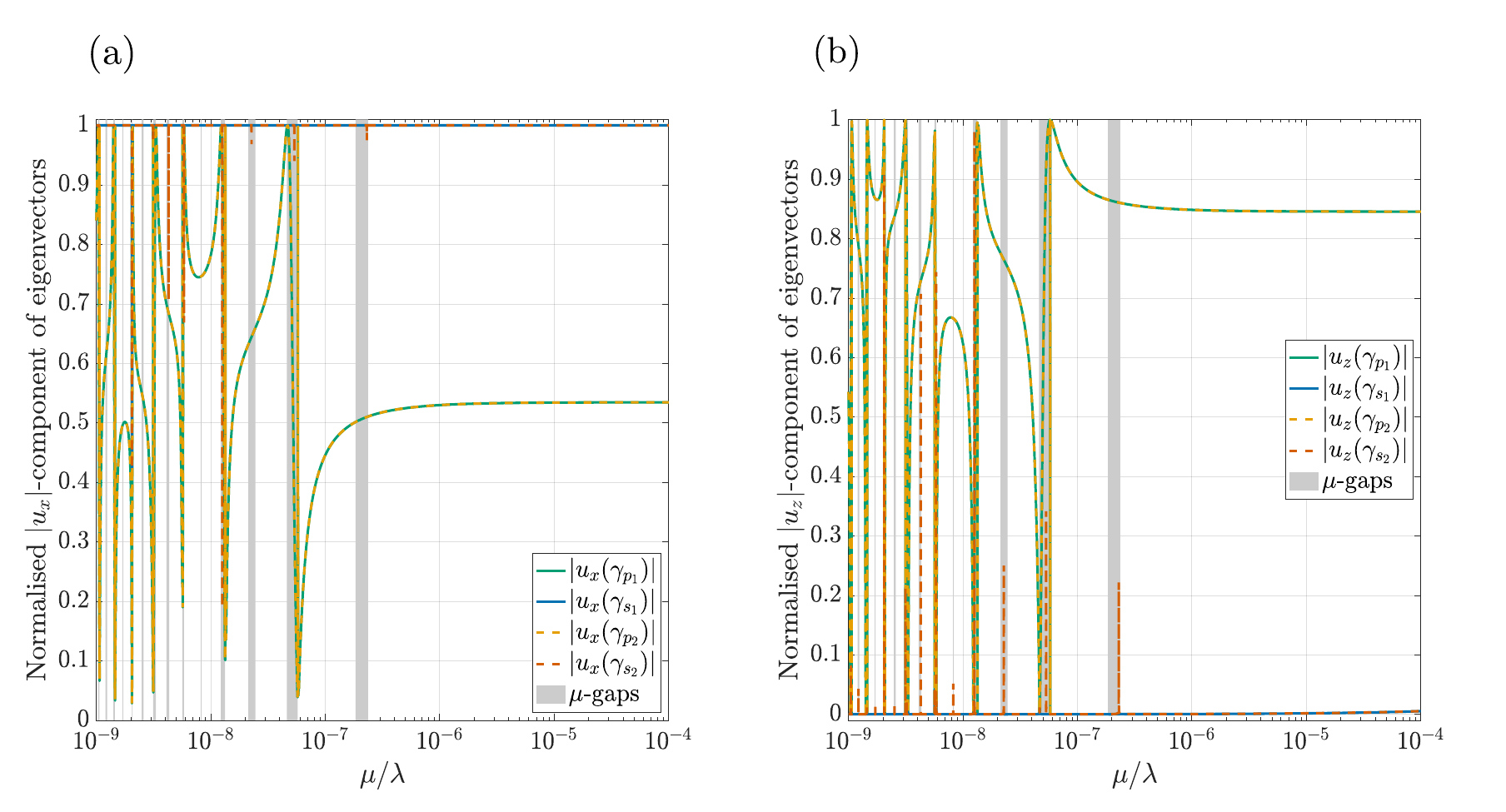}
    \caption{Absolute value of the displacement in the $x$- (a) and $z$-directions (b) associated with both compressional and shear waves against $\mu/\lambda$.}
    \label{Fig_elas_tm_egvec_mu}
\end{figure}

Figure \ref{Fig_elas_tm_egvec_mu} shows the $|u_x|$ (a) and $|u_z|$ (b) components of the eigenvectors as functions of the shear modulus.
The normalisation for these plots is identical to the one described for Figure \ref{Fig_elas_tm_egvec_freq}.
As in Figure \ref{Fig_elas_tm_eigs_mu}, it can be seen that, around the $\mu$-gaps where $\gamma_s$ is positive, the eigenvectors tend to coalesce, implying there are either EPs or ACs near the edges of these $\mu$-gaps.
Additionally, these plots demonstrate that the shear branch is fully shear while the polarisation of the compressional branch can become heavily mixed around the $\mu$-gaps.
This behaviour is analysed in more detail in Section \ref{Sec_IIb}.
Besides, it will become significant in Section \ref{Sec_III} when calculating the effective density of the medium.

\subsection{Bloch wavenumber} \label{Sec_IIa3}

%
To further demonstrate the behaviour of the wave modes in the material, the Bloch wavenumber is now derived and examined. 
From \eqref{Eq_bloch_eig_exp}, the Bloch wavenumbers for compressional and shear waves are calculated as
\begin{equation} \label{Eq_Bloch_wn_gen}
    \mathcal{K}_m = \frac{\ln{\gamma}_m}{ih}, \quad \text{where} \ m = p,s.
\end{equation}
The natural logarithm of the complex-valued eigenvalues $\gamma_m$ is given by \\ ${\ln{\gamma_m} = \ln{|\gamma_m|} + i \left( \varphi_{\gamma_m} + 2 \pi j \right)}$, with $j = 0,1,2,...$.
Here, $\varphi_{\gamma_m}$ corresponds to the phase of the eigenvalue $\gamma_m$.
If both pairs of eigenvalues satisfy that $|\gamma_m| = 1$, $\ln{\gamma_m} = i \varphi_{\gamma_m}$ for $j = 0$, and the wavenumber becomes $\mathcal{K}_m = \varphi_{\gamma_m}/h$.
This corresponds to a purely real wavenumber which, according to \eqref{Eq_bloch_theo}, implies that waves can propagate freely in the medium.
Conversely, in the regions where $|\gamma_m| > 1$ or $|\gamma_m| < 1$, the eigenvalues become real, meaning that $\varphi_{\gamma_m} = 0$ and ${\ln{\gamma_m} = \ln{|\gamma_m|}}$.
In this case, for $j = 0$ the wavenumber is $\mathcal{K}_m = - i \ln{|\gamma_m|}/h$, which is a purely imaginary quantity.
This implies an exponentially decaying wave amplitude in the positive (negative) $z$-direction for positive (negative) real part of the wavenumber. 
Thus, these correspond to non-propagating solutions which are generally referred to as evanescent waves \cite{Blackstock,Anselmet}, further confirming that the $\mu$-intervals where $|\gamma_m| \neq 1$ are analogue to frequency band gaps.

\begin{figure}[h!]
    \centering
    \includegraphics[width=\linewidth,trim={0 0.32cm 0 0},clip]{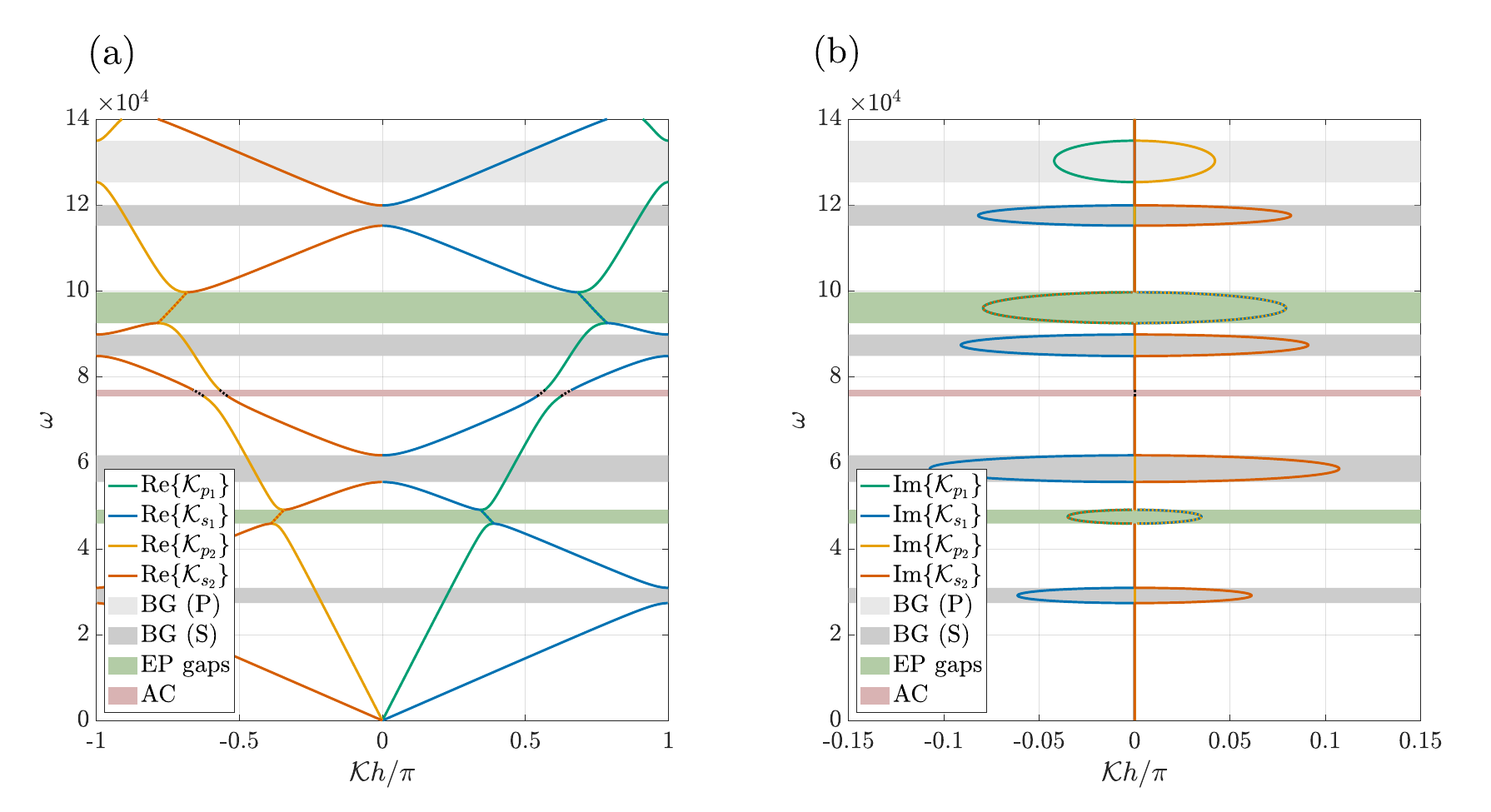}
    \caption{Dispersion diagram, i.e., real part (a) and imaginary (b) parts of the Bloch wavenumber for both compressional and shear waves against frequency. $\mathcal{K}_{s_1}$ and $\mathcal{K}_{s_2}$ are plotted with dotted lines inside the EP gaps.}
    \label{Fig_elas_tm_bloch_freq}
\end{figure}

Figure \ref{Fig_elas_tm_bloch_freq} shows the real (a) and imaginary (b) parts of the Bloch wavenumber against angular frequency $\omega$ calculated from \eqref{Eq_Bloch_wn_gen}.
This corresponds to the dispersion curve of the layered medium.
This allows to clearly identify that the band gaps correspond to frequency ranges where the wavenumbers become purely imaginary.
Regarding the EP gaps, it can be observed that, for these intervals, the wavenumber is now complex.
These also correspond to evanescent waves but now with a non-constant real part of the wavenumber.
Lastly, the behaviour around the avoided crossing is analogous to the one described for the eigenvalues in Figure \ref{Fig_elas_tm_eigs_freq}.

\begin{figure}[h!]
    \centering
    \includegraphics[width=\linewidth,trim={0 0.32cm 0 0},clip]{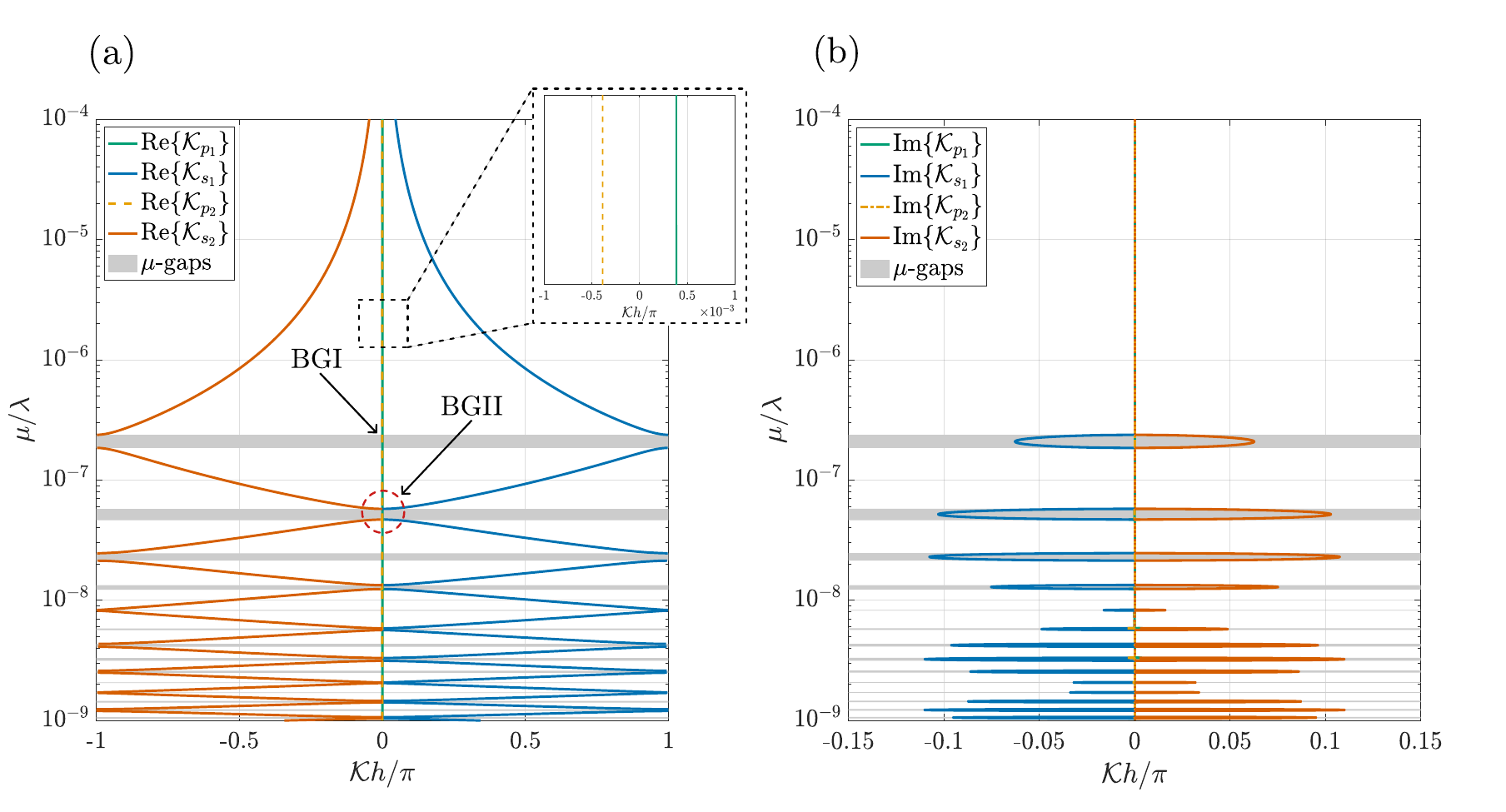}
    \caption{Real part (a) and imaginary (b) parts of the Bloch wavenumber for both compressional and shear waves against $\mu/\lambda$.}
    \label{Fig_elas_tm_bloch_mu}
\end{figure}

As previously described in Figure \ref{Fig_elas_tm_eigs_mu}, the eigenvalues $\gamma_{s}$ and $\gamma_{p}$ become real and approach unity as $\mu$ increases.
In particular, when $\mu$ becomes comparable to $\lambda$, the eigenvalues satisfy that $\gamma_{s} = \gamma_{p} \approx 1$, meaning that their phase is very close to zero and the Bloch wavenumbers become $\mathcal{K}_s (\mu \rightarrow \lambda) \approx \mathcal{K}_p (\mu \rightarrow \lambda) \rightarrow 0$.
The behaviour of $\mathcal{K}_s$ as $\mu \rightarrow 0$ is presented in Figure \ref{Fig_elas_tm_bloch_mu}, which shows the real (a) and imaginary (b) parts of the Bloch wavenumber as functions of the shear modulus.
The real parts of $\mathcal{K}_s$ feature multiple $\mu$-gaps, intervals of $\mu$ where $\mathcal{K}_s$ become purely imaginary.
As $\mu$ decreases, these $\mu$-gaps become narrower and move closer together.
Regarding compressional waves, their wavenumbers remain real, close to zero, and mostly constant for the range displayed in the figure.
A notable exception to this behaviour occurs around certain $\mu$-gaps, where, due to the interaction between modes, the compressional wavenumber can become complex.
This behaviour is explored further in the next section.

\newpage
\subsection{Local behaviour around $\mu$-gaps}
\label{Sec_IIb}

%
This section is devoted to the analysis of the local behaviour of the Bloch wavenumbers around the edges of $\mu$-gaps.
This behaviour is determined by the appearance of EPs and ACs on the right-hand side (larger $\mu$) or left-hand side (smaller $\mu$) of the $\mu$-gap, respectively.
From Figures \ref{Fig_elas_tm_eigs_mu} and \ref{Fig_elas_tm_bloch_mu}, it can be inferred that the $\mu$-gaps that lead to atypical behaviour in the compressional mode occur in an alternating manner.
Firstly, when the real part of the shear-wave eigenvalues approaches $-1$, there is no apparent interaction between shear and compressional modes. 
%
This type of $\mu$-gap is called BGI moving forward.
In contrast, when the real part of the shear eigenvalues approaches $1$, a second type of $\mu$-gap is observed, in which atypical behaviour is identified for compressional waves due to mode interaction with shear waves.
As it will be illustrated in this section, this interaction can take two distinct forms.
From now on, these are referred to as BGIIa and BGIIb.
In the following figures, examples of the wavenumbers, i.e., dispersion diagrams, at each side of both of these types of $\mu$-gaps are presented.

\begin{figure}[h!]
    \centering
    \includegraphics[width=\linewidth,trim={0 0.24cm 0 0},clip]{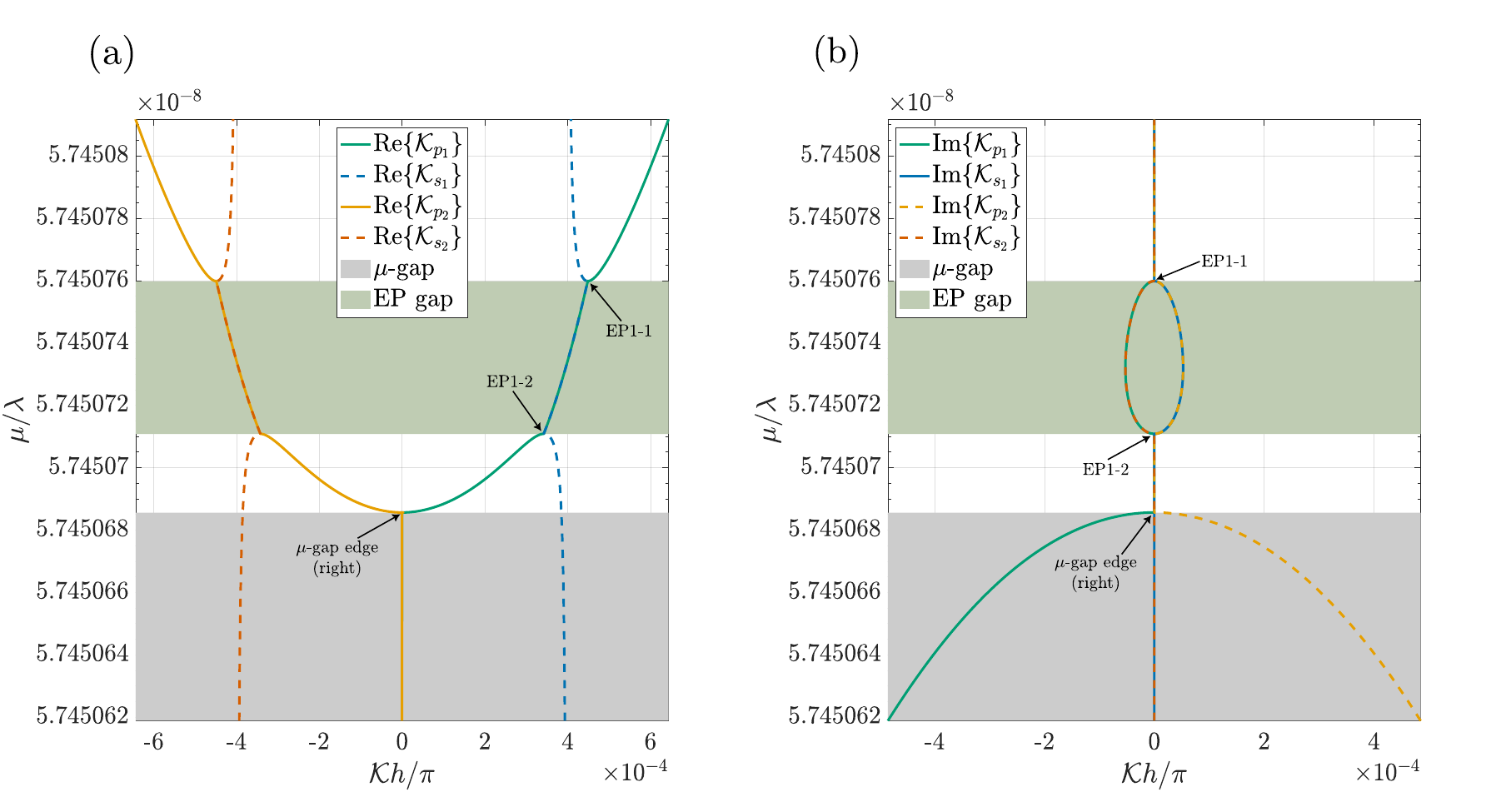}
    \caption{Dispersion curves for both compressional and shear waves against $\mu/\lambda$ on the right-hand side edge of BGIIa-type $\mu$-gap.}
    \label{Fig_elas_tm_disp_curve_right_edge_bg2a}
\end{figure}

Figure \ref{Fig_elas_tm_disp_curve_right_edge_bg2a} shows the dispersion diagram for compressional and shear modes around the right-hand edge of a BGIIa-type $\mu$-gap, i.e., the edge located at a larger value of $\mu$.
The main peculiarity observed is the appearance of a `bump' just before the edge of the $\mu$-gap, where the imaginary parts of both wavenumbers become different from zero.
Moreover, the positive and negative real parts for both modes are equal.
Inside the EP gap the Bloch wavenumber turns complex.
This implies that, due to mode conversion at the boundaries, there exists a range of values of $\mu$ for which the waves become evanescent and thus cannot propagate in the material.
After this EP gap, the shear mode enters a $\mu$-gap, while the compressional waves can propagate freely in the material.
It is important to note that, contrary to the case of the EP gaps in frequency illustrated in Figures \ref{Fig_elas_tm_eigs_freq}, \ref{Fig_elas_tm_egvec_freq}, and \ref{Fig_elas_tm_bloch_mu}, the range of values of $\mu$ where the EP gap appears is considerably smaller than the width of the following $\mu$-gap.
Therefore, the described behaviour is highly localised, and, for the purposes of deriving effective properties, it can be considered as a slight extension of the usual $\mu$-gap.
Even more, analysis of the eigenvectors shows that in the EP gap, for the compressional mode, the longitudinal displacement component no longer dominates, and the wave becomes more shear polarised.
Interestingly, this behaviour is inverted for the stress components, meaning the value of $|\sigma_{zz}|$ for the shear mode increases near the EPs while $|\sigma_{xz}|$ decreases.
Thus, the modes become heavily hybridised and they cannot be unmistakably labelled as either compressional or shear. 

\begin{figure}[h!]
    \centering
    \includegraphics[width=\linewidth,trim={0 0.32cm 0 0},clip]{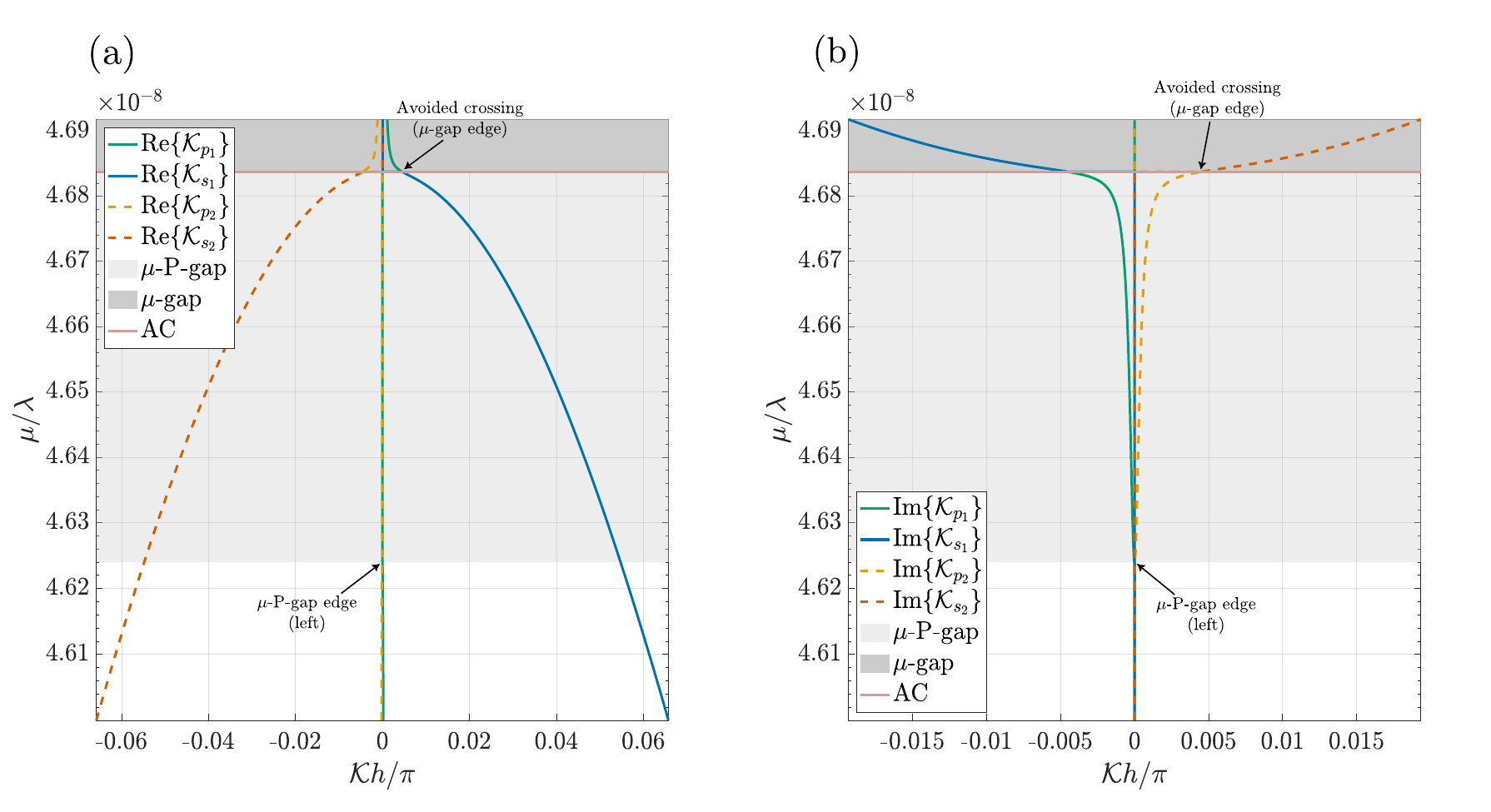}
    \caption{Dispersion curves for both compressional and shear waves against $\mu/\lambda$ on the left-hand side edge of BGIIa-type $\mu$-gap.}
    \label{Fig_elas_tm_disp_curve_left_edge_bg2a}
\end{figure}

The dispersion curves at the left-hand edge of the BGIIa-type $\mu$-gap are shown in Figure \ref{Fig_elas_tm_disp_curve_left_edge_bg2a}, which corresponds to the lower value of $\mu$.
The atypical behaviour observed in this case is determined by the presence of an avoided crossing right before the edge of the $\mu$-gap, and located inside of it.
This implies that, immediately before the $\mu$-gap edge, the labelling of each branch as compressional or shear has to switch.
This choice leads to a discontinuity at said edge.
However, the branches are labelled in such a way that their behaviour outside the $\mu$-gap matches their physical interpretation.
That is, the compressional and shear Bloch wavenumbers, after the avoided crossing, behave as in Figure \ref{Fig_elas_tm_bloch_mu}, with the compressional mode remaining relatively constant and the shear mode oscillating.
Regarding the behaviour right after the AC, the real part of the compressional wavenumber goes to zero and as such it becomes purely imaginary.
This behaviour corresponds to a $\mu$-gap for compressional waves, and is marked as a lighter grey region in the figure.
These are referred to as $\mu$-P-gaps from now on to differentiate them from the $\mu$-gap for shear waves.
Therefore, it can be concluded that, due to the interaction between modes, there is a $\mu$-interval just next to the BGIIa-type $\mu$-gaps where compressional waves cannot propagate in the material.
The eigenvectors also suggest that there is no clear distinction between wave modes in the interval between the avoided crossing and the edge of the $\mu$-P-gap.

\begin{figure}[h!]
    \centering
    \includegraphics[width=\linewidth,trim={0 0.32cm 0 0},clip]{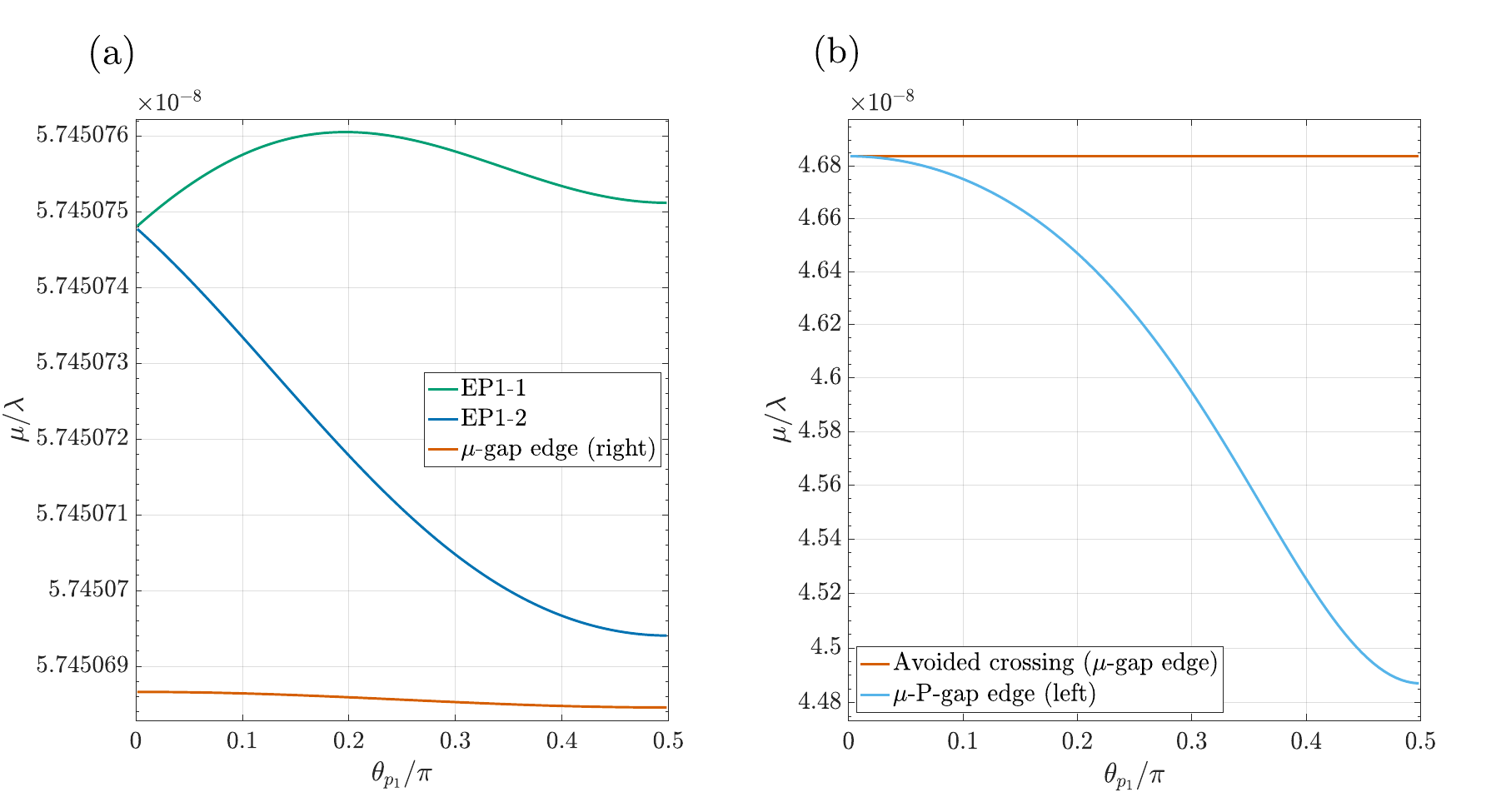}
    \caption{Location in $\mu$-space of EPs, right-hand side of $\mu$-gap (a), avoided crossing, and left-hand side of $\mu$-P-gap (b) as a function of the angle $\theta_{p_1}$ for BGIIa-type $\mu$-gap.}
    \label{Fig_elas_tm_right_edge_bg2a_var_theta}
\end{figure}

The atypical behaviour described in Figures \ref{Fig_elas_tm_disp_curve_right_edge_bg2a} and \ref{Fig_elas_tm_disp_curve_left_edge_bg2a} is also dependent on the propagation angles. 
Figure \ref{Fig_elas_tm_right_edge_bg2a_var_theta}a shows the trajectories of both EP1-1 and EP1-2, as labelled in Figure \ref{Fig_elas_tm_disp_curve_right_edge_bg2a}a, as functions of the propagation angle of compressional waves in the first layer $\theta_{p_1}$.
This plot clearly demonstrates the strong dependency between the exceptional points and $\theta_{p_1}$.
Moreover, it allows to identify that the $\mu$-position of EP1-2 decreases as $\theta_{p_1}$ increases, while EP1-1 is located at the largest possible value of $\mu$ at a propagation angle of around $\pi/5$.
Additionally, it can be observed that the $\mu$-position of the right-hand edge of the $\mu$-gap also varies slightly with the propagation angle.
Nonetheless, this variation is considerably less pronounced compared to that of the EPs, and both edges of the $\mu$-gap are virtually independent of $\theta_{p_1}$. 
This is explained by the fact that, for the small values of $\mu$ where $\mu$-gaps manifest, $c_{s_q} \ll c_{p_1}$ for $q=1,2$ for the first and second layers, respectively.
From \eqref{Eq_snells_law}, it follows that $\sin{\theta_{s_q}} = c_{s_q} \sin{\theta_{p_1}}/c_{p_1} \ll 1$ independently of $\theta_{p_1}$.
Therefore, $\sin{\theta_{s_q}}$ is always close to zero and thus the variation of the position of the $\mu$-gaps edges is negligible. 

An identical examination can be performed for the location of the AC in Figure \ref{Fig_elas_tm_disp_curve_left_edge_bg2a}.
This is shown in Figure \ref{Fig_elas_tm_right_edge_bg2a_var_theta}b, together with the location of the $\mu$-P-gap edge. 
The avoided crossing is largely independent of the propagation angle.
Since the left edge of the $\mu$-gap is determined by the AC, it is also independent of $\theta_{p_1}$.
On the other hand, the edge of the $\mu$-P-gap varies noticeably as $\theta_{p_1}$ increases, with its maximum width being reached when $\theta_{p_1} \rightarrow \pi/2$.
This analysis and the one carried out for the previous figure become relevant in Section \ref{Sec_III} when deriving the effective mass density of the medium as a function of the shear modulus.

\begin{figure}[h!]
    \centering
    \includegraphics[width=\linewidth,trim={0 0.32cm 0 0},clip]{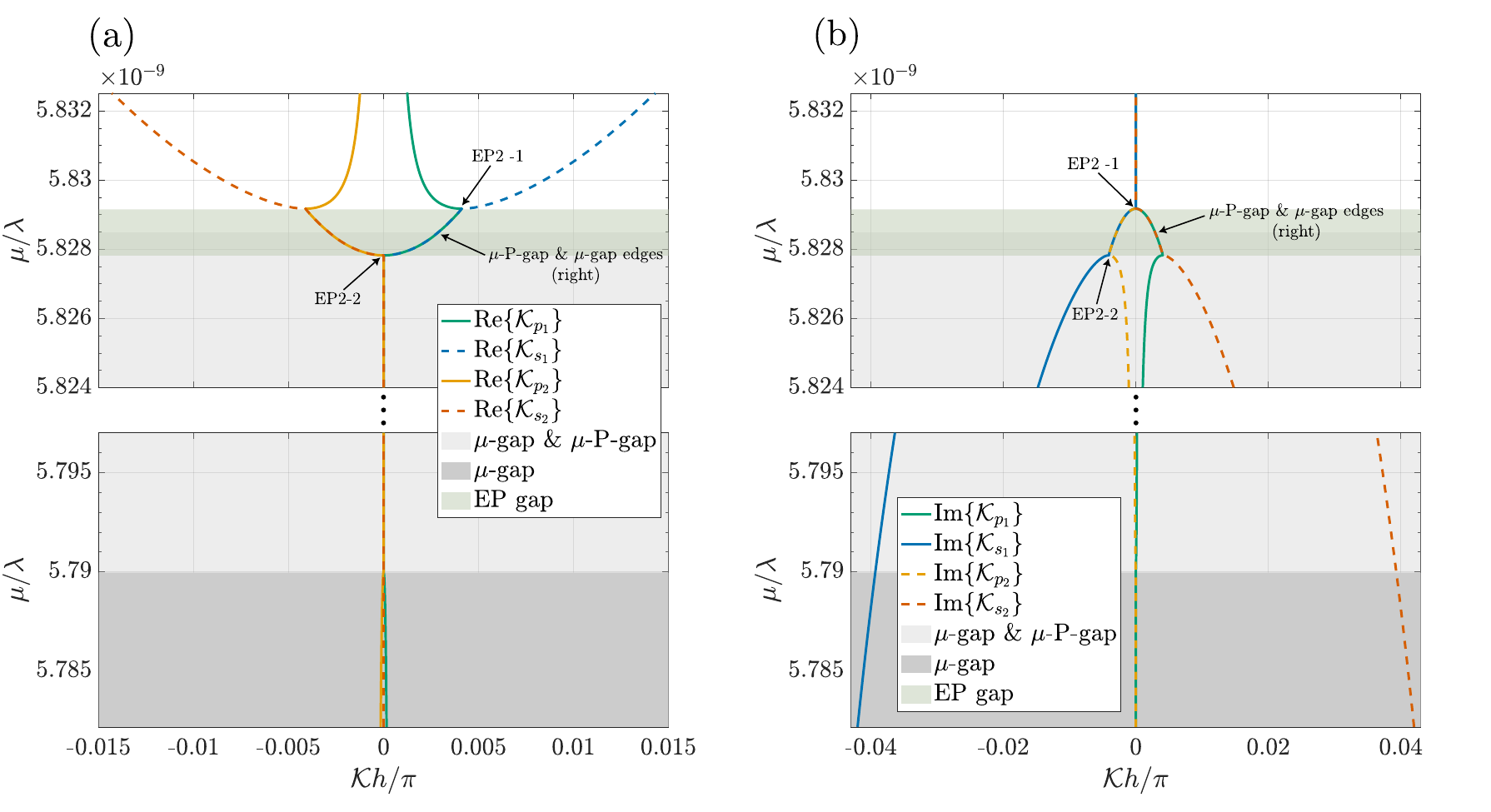}
    \caption{Dispersion curves for both compressional and shear waves against $\mu/\lambda$ on the right-hand side edge of BGIIb-type $\mu$-gap.}
    \label{Fig_elas_tm_disp_curve_right_edge_bg2b}
\end{figure}

\begin{figure}[h!]
    \centering
    \includegraphics[width=\linewidth,trim={0 0.32cm 0 0},clip]{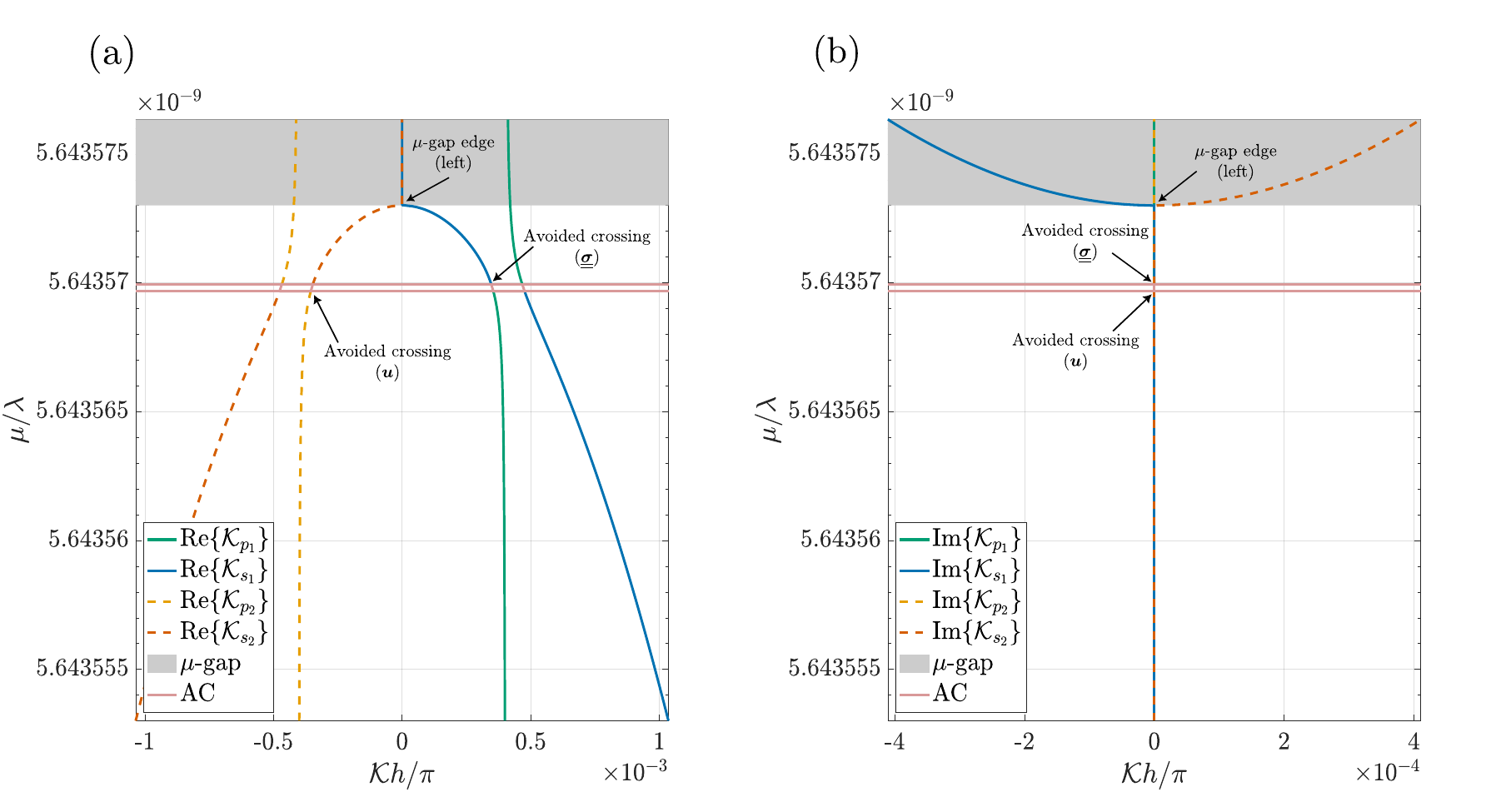}
    \caption{Dispersion curves for both compressional and shear waves against $\mu/\lambda$ on the left-hand side edge of BGIIb-type $\mu$-gap.}
    \label{Fig_elas_tm_disp_curve_left_edge_bg2b}
\end{figure}

As was described at the beginning of this section, there are two types of $\mu$-gaps that lead to atypical behaviour.
The second type of these was previously referred to as BGIIb, and the following figures illustrate the behaviour of the Bloch wavenumber at each side of this type of $\mu$-gap. 
Figure \ref{Fig_elas_tm_disp_curve_right_edge_bg2b} shows the dispersion diagram at the right-hand side edge of a BGIIb-type $\mu$-gap.
As was the case for the BGIIa-type $\mu$-gaps, an EP gap can be identified close to its edge.
The key difference in this case is that one of the exceptional points that delimit the EP gap is located inside the $\mu$-gap.
This implies that the edge of the $\mu$-gap is found inside the EP gap, significantly altering the behaviour of the wavenumbers.
Namely, in between the exceptional point EP2-1 and the $\mu$-gap edge, the dispersion diagram behaves identically to the BGIIa-type $\mu$-gap.
However, between the edge and EP2-2, the absolute value of the imaginary parts keeps increasing while the real parts of both modes quickly tend to zero. 
This in turn leads to a $\mu$-interval after EP2-2 where the wavenumbers for both compressional and shear waves become completely imaginary. 
Therefore, a $\mu$-P-gap which coincides with part of the right side of the $\mu$-gap is identified, forbidding the propagation of both types of waves.
Beyond this interval, only the $\mu$-gap remains and the compressional waves can propagate freely.
As was the case for the BGIIa-type $\mu$-gap, it will be considered that the exceptional point EP2-1, located outside the $\mu$-gap, can be assumed as an extension of the usual $\mu$-gap for the purposes of calculating the effective properties of the material. 
The left-hand side of the BGIIb-type $\mu$-gap is exemplified in the same manner as done for the previous examples.
In that vein, Figure \ref{Fig_elas_tm_disp_curve_left_edge_bg2b} presents the dispersion diagrams for both wave types in the relevant $\mu$-range.
As previously observed for the BGIIa-type $\mu$-gap, the main feature of the left-hand side is the appearance of an AC, which leads to a switch in the labelling of the branches.
Nevertheless, the current $\mu$-gap differs from the one in Figure \ref{Fig_elas_tm_disp_curve_left_edge_bg2a} in that the AC occurs outside of the $\mu$-gap.
This implies that the $\mu$-P-gap observed in the BGIIa-type $\mu$-gaps is prevented from occurring in this case.
After the avoided crossing, for smaller values of $\mu$, the behaviour carries on conventionally for a pass band. 
Additionally, as demonstrated in Figure \ref{Fig_elas_tm_egvec_freq}, the crossings of the components of the eigenvectors do not occur at the same values of $\mu$.
As in that previous example, the stress components intersect at a larger value of $\mu$ than the displacement components.
The crossings are labelled in Figure \ref{Fig_elas_tm_disp_curve_left_edge_bg2b} as avoided crossing ($\tensor{\sigma}$) and avoided crossing ($\vect{u}$).

\begin{figure}[h!]
    \centering
    \includegraphics[width=\linewidth,trim={0 0.32cm 0 0},clip]{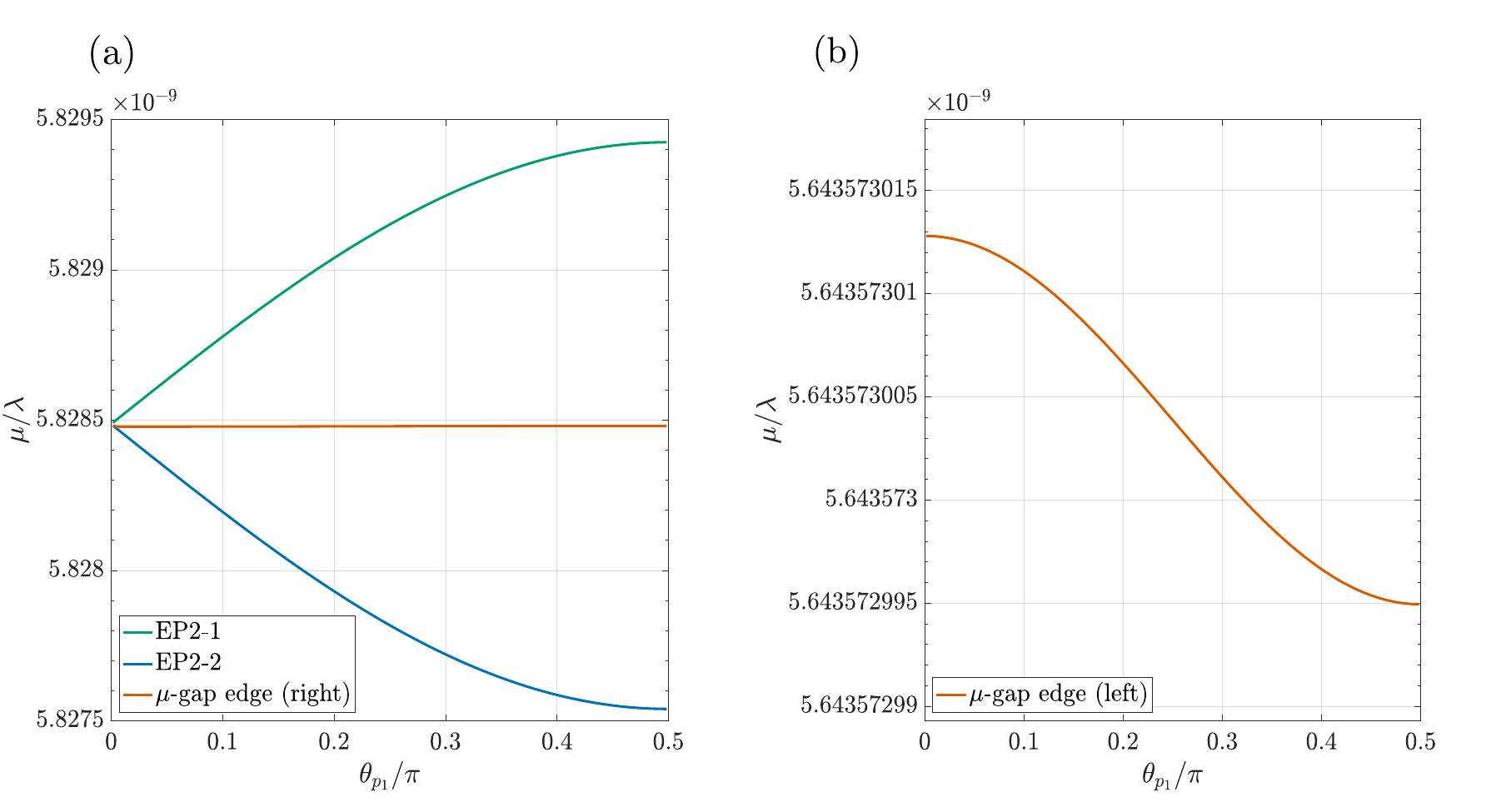}
    \caption{Location in $\mu$-space of EPs, right-hand side of $\mu$-gap (a), and left-hand side of $\mu$-gap (b) as a function of the angle $\theta_{p_1}$ for BGIIb-type $\mu$-gap.}
    \label{Fig_elas_tm_right_edge_bg2b_var_theta}
\end{figure}

Lastly, Figure \ref{Fig_elas_tm_right_edge_bg2b_var_theta}a demonstrates that, as was the case for the BGIIa-type $\mu$-gaps, the EPs are highly dependent on the angle of propagation.
Particularly, as $\theta_{p_1}$ tends to $\pi/2$, EP2-1 moves to a larger value of $\mu$, while EP2-2 does so to a smaller value, thus increasing the width of the EP gap.
On the contrary, the trajectory of the edge of the $\mu$-gap as the propagation angle is varied is largely constant.
This is further demonstrated in Figure \ref{Fig_elas_tm_right_edge_bg2b_var_theta}b, where the variation of the $\mu$-position of the left-hand edge of the $\mu$-gap is shown.
It can be observed that, even though there is some fluctuation in the position of the edge, this is of the order of $O(\mu/\lambda) = 10^{-17}$, eight orders lower than the $\mu/\lambda$ location of the $\mu$-gap, and thus it can be considered negligible. 
The behaviour of the AC when varying the angle of propagation is identical to the one described in Figure \ref{Fig_elas_tm_right_edge_bg2a_var_theta}b, and thus is omitted from the plot.
As mentioned previously, the $\theta_{p_1}$-dependence of the EPs becomes relevant in the following section when calculating the effective density of the medium.

\clearpage
\section{Effective mass density}
\label{Sec_III}

%
Having obtained the Bloch wavenumbers for the layered medium, it is now possible to compute a $\mu$-dependent effective mass density $\tensor{\rho_{\textnormal{eff}}}$.
The tensorial notation is employed to indicate that the effective density can be anisotropic in nature, where the tensor is given by \eqref{Eq_density_tensor} while replacing $\rho_{0x}$ by $\rho_{x\textnormal{eff}}$ and $\rho_{0z}$ by $\rho_{z\textnormal{eff}}$.
The two-layered medium exemplified in previous sections is considered.
To derive its effective density, the following methodology is employed.
A Bloch wavevector is considered, denoted as $\vect{\mathsf{K}}_m = \mathcal{K}_m^t (\sin{\theta_{m_\textnormal{eff}}},\cos{\theta_{m_\textnormal{eff}}})$, where $\mathcal{K}_m^t = |\vect{\mathsf{K}}_m|$ and $\theta_{m_\textnormal{eff}}$ is defined as an effective propagation angle in the material, recalling that $m = p,s$.
The Bloch wavenumber given in \eqref{Eq_Bloch_wn_gen} corresponds to the $z$-component of this wavevector, and thus $\mathcal{K}_m = \mathcal{K}_m^t \cos{\theta_{m_\textnormal{eff}}}$.
Then, an effective phase speed of the layered medium is calculated as
\begin{equation} \label{Eq_c_eff_bloch}
    c_{m_\textnormal{eff}} = \frac{\omega}{\mathcal{K}_m^t} = \frac{\omega}{\mathcal{K}_m} \cos{\theta_{m_\textnormal{eff}}}.
\end{equation}
The effective angle can be calculated from Snell's law in \eqref{Eq_snells_law} as $\sin{\theta_{m_\textnormal{eff}}}/c_{m_\textnormal{eff}} = \sin{\theta_{p_1}}/c_{p_1}$, where the choice of $\sin{\theta_{p_1}}/c_{p_1}$ instead of any other of the terms in \eqref{Eq_snells_law} is arbitrary.
In the $\mu$-intervals where $\mathcal{K}_m$ is real, the phase speed is calculated as
\begin{equation} \label{Eq_c_eff_bloch_comp}
    c_{m_\textnormal{eff}} = \frac{\omega h \cos{\theta_{m_\textnormal{eff}}}}{\varphi_{\gamma_m}},
\end{equation}
where $\varphi_{\gamma_m}$ is the phase of the eigenvalue $\gamma_m$.
The effective angle is then calculated by combining Snell's law with \eqref{Eq_c_eff_bloch_comp} as
\begin{equation} \label{Eq_eff_angle_bloch_comp}
    \tan{\theta_{m_\textnormal{eff}}} = \frac{\omega}{\mathcal{K}_m} \frac{\sin{\theta_{p_1}}}{c_{p_1}}= \frac{\omega h}{\varphi_{\gamma_m}} \frac{\sin{\theta_{p_1}}}{c_{p_1}}.
\end{equation}
To determine the anisotropic effective mass density, an equivalent medium characterised by an unknown anisotropic mass density components $\rho_x$ and $\rho_z$ is considered, defining an unknown phase speed $c$.
The effective density is then obtained by choosing $\rho_x$ and $\rho_z$ such that the phase speed of the equivalent medium matches the effective phase speed calculated from the Bloch wavenumber, i.e., $c_\textnormal{eff} = c$.
Note that this procedure is also valid for acoustics when using the acoustic Bloch wavenumber in \eqref{Eq_c_eff_bloch_comp}.

To simplify the problem and keep the analysis focused on the effective mass density, the following assumptions are considered.
First, it is supposed that wave propagation in the material takes place in the low-frequency regime.
This entails that the wavelength of the waves propagating in the medium is much larger than its characteristic size, i.e., the width of the unit cell.
This leads to $k_1 a \ll 1$ and $k_2 b \ll 1$, where $k_1$ and $k_2$ are the wavenumbers in layers 1 and 2, and $a$ and $b = h -a$ correspond to their thicknesses as given in Figure \ref{Fig_acous_tm_2_layers}.
As a result, the low-frequency approximations for the eigenvalues in acoustics and elasticity shown in Appendix \ref{Append_C} can be taken.
Secondly, it is considered that the layers are isotropic in terms of elasticity constants, and share the same Lam\'{e} parameters, i.e., $\lambda_1 = \lambda_2 = \lambda$ and $\mu_1 = \mu_2 = \mu$, and bulk moduli in the case of acoustics, meaning $K_1 = K_2 = K$.
Considering different elastic constants between layers would lead to an effective medium with a transversely isotropic effective elasticity tensor for layered media \cite{Postma55,Backus62}.
Nonetheless, since the low-frequency effective description is unique, considering either identical or different elastic constants leads to equal results for the effective density \cite{Nunez25}.

\subsection{Acoustics} \label{SecIII_acous}

%
\subsubsection{Effective density in the $z$-direction}

To derive the effective density in the $z$-direction, i.e., $\rho_{z \textnormal{eff}}$, the incident angle is set to zero, leading to $c_\textnormal{eff} = \omega h/\varphi_\gamma$ from \eqref{Eq_c_eff_bloch_comp}.
The general equation for the acoustic phase speed in media with anisotropic mass density in 2D is given in \eqref{Eq_acous_aniso_speed}.
Considering normal incidence, the phase speed reduces to $c = \sqrt{K/\rho_z}$.
Here, $K$ is a given number and thus it is necessary to find the value of $\rho_z$ for which $c = c_{\textnormal{eff}}$.
This leads to
\begin{equation} \label{Eq_acous_eff_rho_z}
    \rho_z = \frac{K \varphi_\gamma^2}{(\omega h)^2}.
\end{equation}
Now remains to calculate the phase $\varphi_\gamma$ of the eigenvalues of the acoustic transfer matrix.
From \eqref{Eq_acous_eig_comp}, the dominant terms in the low-frequency approximation for the acoustic eigenvalues are those that contain linear terms of the wavenumbers $k_1$ and $k_2$.
Consequently, the low-frequency eigenvalues are given by
\begin{equation}
    \gamma \approx 1 + i \sqrt{(k_1 a)^2 + k_1 a k_2 b \left( \frac{r_1}{r_2} + 
     \frac{r_2}{r_1}\right) + (k_2 b)^2},
\end{equation}
where the positive sign is selected since it represents forward-propagating waves.
Thus $\mathrm{Re}\{\gamma\} \approx 1$ and
\begin{equation} \label{Eq_approx_tan_eig}
    \tan{\varphi_\gamma} = \frac{\mathrm{Im}\{\gamma\}}{\mathrm{Re}\{\gamma\}} \approx \mathrm{Im}\{\gamma\} = \sqrt{(k_1 a)^2 + k_1 a k_2 b \left( \frac{r_1}{r_2} + 
      \frac{r_2}{r_1}\right) + (k_2 b)^2}.
\end{equation}
Considering that wave propagation occurs in the low-frequency regime, the approximation %
\\
${\tan{\varphi_\gamma} \approx \varphi_\gamma}$ is considered.
Replacing $k_1 = \omega \sqrt{\rho_1/K}$, $k_2 = \omega \sqrt{\rho_2/K}$, and $r_1 = \sqrt{K\rho_1}$ and ${r_2 = \sqrt{K\rho_2}}$ for normal incidence, leads to the phase of the eigenvalue being expressed as
\begin{equation}
    \varphi_\gamma = \sqrt{\frac{\omega^2 h (a \rho_1 + b \rho_2)}{K}}.
\end{equation}
Replacing this equation into \eqref{Eq_acous_eff_rho_z} and introducing $\varphi = a/h$ leads to
\begin{equation} \label{Eq_eff_rho_z_acous}
    \rho_z = \varphi \rho_1 + (1 - \varphi) \rho_2 = \rho_{z \textnormal{eff}}.
\end{equation}
Consequently, the effective density in the $z$-direction for a infinitely periodic layered medium is shown to correspond to the arithmetic average of the densities of each of the layers.

\subsubsection{Effective density in the $x$-direction}
In this case, it is necessary to consider oblique incidence so that $\rho_x$ is present in the equation for the acoustic phase speed.
As before, to find the effective density, the effective phase speed is set to be equal to the anisotropic phase speed.
The main difference now is that, since $\theta_i \neq 0$, the effective propagation angle must be considered in the calculation of the anisotropic phase speed from \eqref{Eq_acous_aniso_speed}, meaning that the equation to solve is now $c_\textnormal{eff} = c(\theta_{\textnormal{eff}})$.
Doing this and rewriting the equation so that $\rho_x$ appears in only one side of the equivalence leads to
\begin{equation} \label{Eq_acous_eff_dens_rho_x_gen}
    \frac{\sin^2{\theta_1}}{\rho_x} = \frac{c_1^2}{K} - \left( \frac{\varphi_\gamma}{\omega h} \right)^2 \frac{c_1^2}{\rho_z},
\end{equation}
where \eqref{Eq_eff_angle_bloch_comp} was employed to eliminate the dependence on $\theta_{\textnormal{eff}}$.
Following \eqref{Eq_approx_tan_eig}, the phase of the eigenvalues is given solely by their imaginary part as
\begin{equation}
    \varphi_\gamma \approx  \sqrt{(k_1 a \cos{\theta_1})^2 + k_1 a \cos{\theta_1} k_2 b \cos{\theta_2} \left( \frac{r_1}{r_2} + 
     \frac{r_2}{r_1}\right) + (k_2 b \cos{\theta_2})^2}.
\end{equation}
Following the same procedure as for $\rho_z$, by replacing the expression for the wavenumbers and acoustic impedances, the eigenvalue phase can be expressed in terms of the individual densities of the layers, i.e.,
\begin{equation}
    \varphi_\gamma \approx \sqrt{\frac{\omega^2 (a \rho_1 + b\rho_2) (a \cos^2{\theta_1} + b \cos^2{\theta_2})}{K}}.
\end{equation}
This expression for the phase can be directly introduced into \eqref{Eq_acous_eff_dens_rho_x_gen}, which leads to a general equation relating the densities in the $x$- and $z$-directions, namely
\begin{equation}
    \left( \frac{1}{\rho_x} - \frac{1}{\rho_g} \right) \sin^2{\theta} + \left( \frac{1}{\rho_z} - \frac{1}{\rho_a} \right) \cos^2{\theta} = 0,
\end{equation}
where $\rho_a = \varphi \rho_1 + (1 - \varphi)$ and $\rho_g^{-1} = \varphi/\rho_1 + (1 - \varphi)/\rho_2$.
Since from \eqref{Eq_eff_rho_z_acous} it is already known that $\rho_z = \rho_a$, this equation reduces to
\begin{equation} \label{Eq_eff_rho_x_acous}
    \frac{1}{\rho_x} = \frac{\phi}{\rho_1} + \frac{1-\phi}{\rho_2} = \frac{1}{\rho_{x \textnormal{eff}}}.
\end{equation}

From \eqref{Eq_eff_rho_z_acous} and \eqref{Eq_eff_rho_x_acous}, it is clear that the effective density in acoustics is anisotropic.
Moreover, the method employed allowed to correctly obtain the classic results given in literature for the effective density of a layered medium when considering acoustic waves.

\subsection{Elasticity}

%
The analysis is now turned towards the derivation of the effective density in elastodynamics.
The main difficulty that arises in this case comes from the complexity of the expressions for the eigenvalues.
However, by assuming normal incidence, the transfer matrix is significantly simplified, as explained in Section \ref{Sec_IIa}.
The low-frequency eigenvalues for each mode are shown in Appendix \ref{Append_C}.
Then, to find the effective densities in the $x$- and $z$-directions, the effective phase speed is compared with the phase speeds of compressional and shear waves in an elastic medium with anisotropic density as given in \eqref{Eq_elasticity_aniso_p_wave_speed} and \eqref{Eq_elasticity_aniso_s_wave_speed}.

\subsubsection{Effective density in the $z$-direction} \label{SecIII_elas_rho_z}

Following the same procedure as in Section \ref{SecIII_acous}, setting $\theta_i = 0$ leads to $c_{p_\textnormal{eff}} = \omega h/\varphi_{\gamma_p}$.
Under the normal incidence assumption, from \eqref{Eq_elasticity_aniso_p_wave_speed} the phase speed for compressional waves in the effective medium with anisotropic mass density is given by $c_p = \sqrt{(\lambda + 2\mu)/\rho_z}$.
As for acoustics, the effective density is found by comparing the known effective phase speed $c_{p_\textnormal{eff}}$ to the anisotropic phase speed with the unknown density in the $z$-direction.
This leads to
\begin{equation} \label{Eq_elas_eff_dens_z_dir}
    \rho_z = \left( \frac{\varphi_{\gamma_p}}{\omega h} \right)^2 (\lambda + 2\mu),
\end{equation}
From \eqref{Eq_eig_m}, by neglecting higher-order terms, it is obtained that $\mathrm{Re}\{\gamma_p\} \approx 1$ and thus \\
${\varphi_{\gamma_p} \approx \mathrm{Im}\{\gamma_p\}}$, where
\begin{equation}
    \mathrm{Im}\{\gamma_p\} \approx \sqrt{(k_{p_1} a)^2 + (k_{p_2} b)^2 + k_{p_1} a k_{p_2} b \left( \frac{k_{p_1}}{k_{p_2}} + \frac{k_{p_2}}{k_{p_1}} \right)}.
\end{equation}
Importantly, this approximation imposes no restriction on the shear modulus. 
The phase of the eigenvalues can be rewritten in terms of the mass densities of the layers $\rho_1$ and $\rho_2$ as
\begin{equation}
    \varphi_{\gamma_p} = \sqrt{\frac{\omega^2 h (a \rho_1 + b \rho_2)}{\lambda + 2\mu}}.
\end{equation}
By replacing this equation into \eqref{Eq_elas_eff_dens_z_dir}, the density in the $z$-direction is found to be
\begin{equation} \label{Eq_eff_rho_z_elas}
    \rho_z = \phi \rho_1 + (1 - \phi) \rho_2 = \rho_{z \textnormal{eff}}.
\end{equation}

This result is consistent with the literature (see, e.g., \cite{Postma55}), and demonstrates that the effective densities perpendicular to the layers are identical for both acoustics and elasticity.

\subsubsection{Effective density in the $x$-direction} \label{SecIII_elas_rho_x}

For the effective density in the $x$-direction, it is possible to derive an expression by considering normal incidence.
However, this requires that $\rho_{x \textnormal{eff}}$ be derived from the effective phase speed for shear waves, since the shear phase speed is solely dependent in the density in this direction for normal incidence.
This is demonstrated by setting $\theta_i = 0$ in \eqref{Eq_elasticity_aniso_s_wave_speed}, leading to $c_s = \sqrt{\mu/\rho_x}$.
Since $c_{s_\textnormal{eff}} = \omega h /\varphi_{\gamma_s}$, the density in the $x$-direction must be given by $\rho_x =  \mu ( \varphi_{\gamma_s}/\omega h)^2$.
Following a similar reasoning as for $\rho_{z \textnormal{eff}}$, it is found that
\begin{equation} \label{Eq_phase_eig_rho_x_elas}
    \varphi_{\gamma_s} \approx \sqrt{(k_{s_1} a)^2 + (k_{s_2} b)^2 + k_{s_1} a k_{s_2} b \left( \frac{k_{s_1}}{k_{s_2}} + \frac{k_{s_2}}{k_{s_1}} \right)},
\end{equation}
which is equivalent to $\varphi_{\gamma_p}$ but replacing $k_p$ by the wavenumber for shear waves.
Notably, in this case, since $k_{s_1}$ and $k_{s_2}$ are directly dependent on the shear modulus, \eqref{Eq_phase_eig_rho_x_elas} is only valid for values of $\mu$ that satisfy $k_{s_1} a \ll 1$ and $k_{s_2} b \ll 1$.
Therefore, considering the same assumptions as before, the phase is obtained as a function of $\rho_1$ and $\rho_2$ as
\begin{equation}
    \varphi_{\gamma_p} = \sqrt{\frac{\omega^2 h (a \rho_1 + b \rho_2)}{\mu}},
\end{equation}
and thus the effective density must be given by 
\begin{equation} \label{Eq_eff_rho_x_elas}
    \rho_x = \phi \rho_1 + (1 - \phi) \rho_2 = \rho_{x \textnormal{eff}}.
\end{equation}

The results presented in \eqref{Eq_eff_rho_z_elas} and \eqref{Eq_eff_rho_x_elas} confirm that $\rho_{z \textnormal{eff}} = \rho_{x \textnormal{eff}}$, thus demonstrating that, for elasticity, the effective density is isotropic.
Moreover, this corresponds to the standard result in this regime, verifying that the Bloch wavenumber approach allows to obtain the expected effective density.

\subsection{Transition from elasticity to acoustics}

%
Having shown that it is possible to recover the usual effective mass density for acoustics and elasticity using the Bloch wavenumber as a starting point, the natural next step is to analyse the behaviour of $\rho_x$ and $\rho_z$ when considering the 'acoustic limit'.
For this, as in Section \ref{SecIII_elas_rho_z}, the starting point is the Bloch wavenumber for compressional waves $\mathcal{K}_p$.
This choice is made since the main objective of this work is to show that the elastic effective mass density tends to its acoustic counterpart when $\mu \rightarrow 0$.
In this limit, the compressional waves in elasticity become acoustics waves, while the shear mode disappears. 
This presents the difficulty that $\rho_x$ and $\rho_z$ only appear simultaneously in \eqref{Eq_elasticity_aniso_p_wave_speed} when oblique incidence is considered.
Choosing $\theta_i \neq 0$ has the consequence that mode conversion will take place at the boundaries between layers, and thus the compressional and shear branches will be coupled, leading to the behaviour described in Section \ref{Sec_IIb} as the shear modulus tends to zero.
Therefore, the effects these interactions between shear and compressional modes have on the effective density is also explored in this section.

From \eqref{Eq_eff_rho_z_acous} and \eqref{Eq_eff_rho_z_elas}, it is identified that the effective mass density in the $z$-direction $\rho_{z \textnormal{eff}}$ matches for both acoustics and elasticity.
Moreover, the expression for this effective density in elasticity given in \eqref{Eq_eff_rho_z_elas} was found without imposing any restrictions on the value of $\mu$.
Thus, it is inferred that the $z$-component of the effective density is independent of the shear modulus and remains constant as $\mu \rightarrow 0$.
The case of $\rho_{x \textnormal{eff}}$ is trickier since the limits do not match as seen in \eqref{Eq_eff_rho_x_acous} and \eqref{Eq_eff_rho_x_elas}.
In addition, for a sufficiently small value of $\mu$, particularly if $k_{s_1} a$ and $k_{s_2} b$ are close to one, \eqref{Eq_eff_rho_x_elas} ceases to be valid.
Therefore, there must be a $\mu$-dependent expression for this density that allows to find the transition between these limits. 
To find such an expression, it is considered that the effective density in the $x$-direction $\rho_{x \textnormal{eff}}$ corresponds to the solution of $c_p(\rho_{x \textnormal{eff}},\rho_{z \textnormal{eff}}) = c_{p_\textnormal{eff}}$.
Solving for $\rho_{x \textnormal{eff}}$ leads to
\begin{equation} \label{Eq_eff_dens_eigs}
    \rho_{x \textnormal{eff}} = \frac{c_{p_\textnormal{eff}}^2 \Gamma_{11} \rho_{z \textnormal{eff}} - \Gamma_{11} \Gamma_{22} + \Gamma_{12}^2}{c_{p_\textnormal{eff}}^2 \left(c_{p_\textnormal{eff}}^2 \rho_{z \textnormal{eff}} - \Gamma_{22} \right)},
\end{equation}
where $c_{p_\textnormal{eff}}$ is given by \eqref{Eq_c_eff_bloch} with $m = p$.
Here, it is recalled that $c_{p_\textnormal{eff}}$ is a function of the eigenvalues of the transfer matrix, which in turn are functions of the shear modulus, among other parameters. 
Additionally, $\Gamma_{11}$, $\Gamma_{22}$, and $\Gamma_{12}$ are given by \eqref{Eq_elasticity_aniso_christoffel_matrix_comp_1}, and thus are also functions of $\mu$.

Eq.~\eqref{Eq_eff_dens_eigs} constitutes the first analytical expression for the effective density of an periodic elastic layered medium as a function of the shear modulus $\mu$, to the best of the authors' knowledge. 
Nevertheless, there are some caveats to this equation that are relevant to address.
Firstly, it is direct to identify that there are certain values of $\mu$ that can lead to discontinuities in the effective density.
This occurs when $c_{p_\textnormal{eff}}^2 \rho_{z \textnormal{eff}} = \Gamma_{22}$, which expanded and combined with \eqref{Eq_eff_angle_bloch_comp} leads to an equation for the (positive) wavenumber that leads to the denominator in \eqref{Eq_eff_dens_eigs} becoming zero as
\begin{equation} \label{Eq_K_discont}
    \mathcal{K}_{\textnormal{disc}}(\mu) = \omega \sqrt{\frac{\rho_{z\textnormal{eff}} - \mu \sin^2{\theta_{p_1}}/c_{p_1}^2}{\lambda + 2\mu}} \cdot
\end{equation}
For values of $\mathcal{K}_p < \mathcal{K}_{\textnormal{disc}}$, the effective density is positive, while it becomes negative when the wavenumber is larger than $\mathcal{K}_{\textnormal{disc}}$.
Additionally, for a purely real wavenumber, and considering that $\mathrm{Re}\{\gamma_p\} \approx 1$, this condition translates to
\begin{equation} \label{Eq_gamma_discont}
    \mathrm{Im}\{\gamma_p(\mu)\} = \omega h \sqrt{\frac{\rho_{z\textnormal{eff}} - \mu \sin^2{\theta_{p_1}}/c_{p_1}^2}{\lambda + 2\mu}} \cdot
\end{equation}
As it will be shown in the ensuing paragraphs, for the majority of the $\mu$-domain the effective density remains positive.
However, around the BGII-type $\mu$-gaps, particularly in the ranges shown in Figures \ref{Fig_elas_tm_disp_curve_left_edge_bg2a} and \ref{Fig_elas_tm_disp_curve_right_edge_bg2b}, the imaginary part of the eigenvalues can increase in value, eventually crossing a discontinuity, forcing the effective density to become negative for a narrow $\mu$ interval.

Secondly, the nature of the effective density is entirely dependent on the effective compressional phase speed, which in turn is determined by the compressional Bloch wavenumber.
In pass bands, the wavenumber is completely real and thus the effective density will also be a purely real number. 
Particular attention must be given to the $\mu$-intervals where the wavenumber becomes complex, i.e., $\mu$-P-gaps and EP gaps, detailed in the previous section.
In these gaps, the effective phase speed and effective propagation angle, as calculated from \eqref{Eq_c_eff_bloch} and \eqref{Eq_eff_angle_bloch_comp}, respectively, can become imaginary or complex.
Consequently, \eqref{Eq_eff_dens_eigs} can lead to a complex effective density. 
In essence, the analytical model can be safely employed to calculate $\rho_{x \textnormal{eff}}$ when $\mathcal{K}_p \in \mathbb{R}$, but special care must be taken in the $\mu$-intervals where the wavenumber is not completely real.

It is likewise relevant to analyse the dependence of the effective density on the angle of incidence $\theta_i$, considering it is of interest to obtain an effective mass density that is intrinsic to the medium and thus independent of the incident angle of the waves.
This is achieved by calculating $\rho_{x \textnormal{eff}}$ for several values of $\theta_i$.
Afterwards, the relative error between the values of the effective density for the different propagation angles is calculated as
\begin{equation}
    E(\theta_i) = \left|\frac{\rho_{x \textnormal{eff}} (\pi/3) - \rho_{x \textnormal{eff}} (\theta_i)}{\rho_{x \textnormal{eff}} (\pi/3)}\right|,
\end{equation}
where $\rho_{x \textnormal{eff}} (\theta_i)$ is the effective density for a given angle $\theta_i$, and $\rho_{x \textnormal{eff}} (\pi/3)$ is chosen as the reference value.

\begin{figure}[h!]
    \centering
    \includegraphics[width=\linewidth,trim={0 0 0 0},clip]{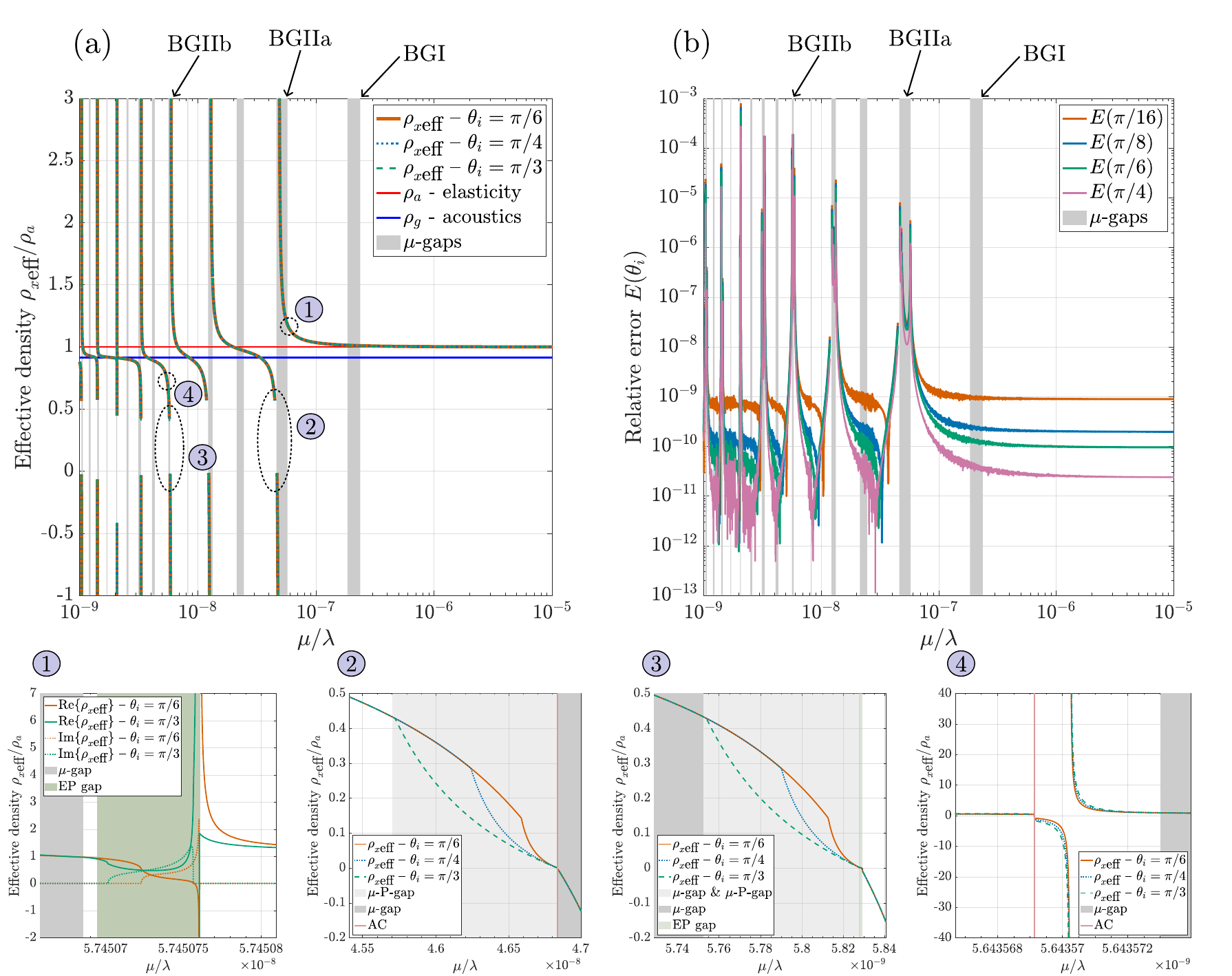}
    \caption{(a) Effective density $\rho_{x \textnormal{eff}}$ as a function of $\mu/\lambda$ when $\mu \rightarrow 0$ (acoustic limit) for different incident angles and (b) relative error for $\rho_{x \textnormal{eff}}$ between the different incident angles.}
    \label{Fig_elas_eff_dens_bloch}
\end{figure}

To illustrate the behaviour of the effective density in the $x$-direction as the shear modulus tends to zero, an example is presented in Figure \ref{Fig_elas_eff_dens_bloch}.
However, before discussing this figure, it is relevant to indicate the choices made regarding the atypical behaviour brought about due to the presence of EPs for the BGII-type $\mu$-gaps. 
As mentioned in the previous paragraph, inside the $\mu$- and EP gaps the effective density can become complex.
Moreover, since the $\mu$-location of the EPs is highly dependent on $\theta_i$, it is expected that the effective density derived from \eqref{Eq_eff_dens_eigs} will also be very sensitive to the incident angle. 
This is exemplified in Insert 1 of Figure \ref{Fig_elas_eff_dens_bloch}, which corresponds to the $\mu$-range shown in Figure \ref{Fig_elas_tm_disp_curve_right_edge_bg2a}, demonstrating an angle-dependent complex effective density inside the EP gap.
For these reasons, it is considered that in these gaps it is not possible to derive an effective description for the layered medium. 
Even further, to ensure that the effective density is completely independent of $\theta_i$, it is necessary to exclude the entire range of $\mu$-locations of the exceptional points as functions of $\theta_i$.
This is achieved by numerically calculating, for every BGII-type $\mu$-gap located in the $\mu$-range of interest, the largest values of EP1-1 and EP2-1, and the smallest values of EP1-2 and EP2-2, both as functions of $\theta_i$, as shown in Figure \ref{Fig_elas_tm_disp_curve_right_edge_bg2a} and \ref{Fig_elas_tm_disp_curve_right_edge_bg2b}.
Afterwards, it is considered that, in the ranges between these values of EP1-1 and EP1-2, and EP2-1 and EP2-2, an effective density cannot be defined. 
The same procedure is applied for $\mu$-P-gaps that appear around the BGII-type $\mu$-gaps, shown in Inserts 2 and 3 of Figure \ref{Fig_elas_eff_dens_bloch} for the BGIIa- and BGIIb-type $\mu$-gaps.
These also correspond to Figures \ref{Fig_elas_tm_disp_curve_left_edge_bg2a} and \ref{Fig_elas_tm_disp_curve_right_edge_bg2b}, respectively.
These figures also verify that the effective density is dependent on the incident angle in these ranges.
Furthermore, even though the $\mu$-location of the avoided crossings is not affected by a change of $\theta_i$, it has been identified that the effective density $\rho_{x \textnormal{eff}}$ around them is dependent on the incident angle, which is displayed in Insert 4 of Figure \ref{Fig_elas_eff_dens_bloch} for the $\mu$-range shown in Figure \ref{Fig_elas_tm_disp_curve_left_edge_bg2b}.
As such, it is considered that the effective properties also cannot be defined around these crossings. 
To summarise, all the gaps in the curves shown in Figure \ref{Fig_elas_eff_dens_bloch}a correspond to $\mu$-intervals where the effective mass is not uniquely defined, owing to its dependence on the propagation angle of the waves.
These gaps occur exclusively at the edges of the BGII-type $\mu$-gaps.

Figure \ref{Fig_elas_eff_dens_bloch} shows the effective density $\rho_{x \textnormal{eff}}$ as a function of the shear modulus $\mu$ plotted for three different angles of incidence.
It can be observed that, as the shear modulus increases and gets closer to $\lambda$, the effective density tends to its elastic counterpart, i.e., the arithmetic average $\rho_a$, confirming that, for these values of $\mu$, wave propagation takes place in the usual elasticity regime.
However, it is not clear uniquely from this figure that the acoustic limit is reached as the shear modulus tends to zero.
This is mainly due to the discontinuities in the effective density that begin to appear as the shear modulus decreases. 
As explained earlier, these occur for values of $\mu$ that satisfy \eqref{Eq_K_discont} and \eqref{Eq_gamma_discont}, and they appear at the left-hand side of the BGIIa-type $\mu$-gaps, and to the right-hand side of the BGIIb-type $\mu$-gaps.
Importantly, these coincide with the $\mu$-P-gaps described in the previous section.
Additionally, $\rho_{x \textnormal{eff}} < 0$ is observed for values of the shear modulus close to a $\mu$-P-gap, while the effective mass density remains positive for all other $\mu$.
Therefore, it is concluded that the existence of $\mu$-P-gaps brought about by mode coupling due to mode conversion at the boundaries leads to the atypical behaviour observed in the figure. 
Accordingly, since the $\mu$-P-gaps and thus the discontinuities in the effective density seem to appear periodically as $\mu$ tends to zero, a definitive outcome for the $\mu \rightarrow 0$ limit cannot be reached only through this example. 
However, it appears that, for the pass band in $\mu$ shown in the figure, by not considering the discontinuities in the effective density, $\rho_{x \textnormal{eff}}$ tends to its acoustic counterpart.
This is explored in the next examples.

Despite this, two important conclusions can be drawn from the current figure.
Firstly, it allows to reach the same conclusion as in \cite{Nunez25}, namely, that the effective density in the elastic regime is \textbf{not} necessarily isotropic, and that considering a small enough value of $\mu$ leads to an anisotropic effective density.
Secondly, the curves in Figure \ref{Fig_elas_eff_dens_bloch} confirm the $\theta_i$-independence of the effective density in the ranges of $\mu$ where \eqref{Eq_eff_dens_eigs} is well-behaved.
From Figure \ref{Fig_elas_eff_dens_bloch}b, it can be observed that, by not considering the $\mu$-ranges where the effective density is ill-defined, the maximum error between $\rho_{x \textnormal{eff}}$ calculated for different incident angles is no more than 0.1\%.
Moreover, the error is clustered around the edges of the BGII-type band gaps.
In the pass bands, the actual error is no larger than $10^{-6}$\%, implying the effective density is largely independent of the angle of incidence.
For comparison, the maximum relative error when considering the band gaps, EP gaps, and avoided crossings, climbs up to almost 100\%, which confirms that it is not possible to define an effective mass density in these gaps.

\subsection{Validation with self-consistent method and acoustic limit}
As a way to validate the results presented in the previous figure, the effective density derived from the Bloch wavenumber is compared with the results of applying a dynamic self-consistent method \cite{yang2003} to a layered medium consisting of a large number of unit cells.
For this, the same procedure employed in \cite{Nunez25} is implemented.
This consists in placing a finite layered medium in between two identical half-spaces (see Figure \ref{Fig_elas_tm_1_layer}) exhibiting anisotropic mass density.
The transfer matrix approach allows to directly consider any given number $n$ of unit cells.
The effective density is then determined as the value of the density of the half-spaces that leads to zero reflection and full transmission across the material.
The procedure is as follows: from a starting guess for $\rho_{0 x}$ a function $\vect{F} = (|\mathscr{R}_p|,|1 - \mathscr{T}_p|)$ is calculated, where $\mathscr{R}_p$ and $\mathscr{T}_p$ are the complex-valued reflection and transmission coefficients for compressional waves, respectively.
Then, by employing a gradient-descent algorithm, the value of $\rho_{0 x}$ that leads to $\vect{F} = \vect{0}$ is found and chosen as the effective density of the layered material, i.e., $\rho_{x \textnormal{eff}} = \rho_{0 x}$.
This is repeated for the whole range of $\mu$ of interest, using the previous result as a starting point for the next value of $\mu$.

\begin{figure}[h!]
    \centering
    \includegraphics[width=0.7\linewidth,trim={0 0.32cm 0 0},clip]{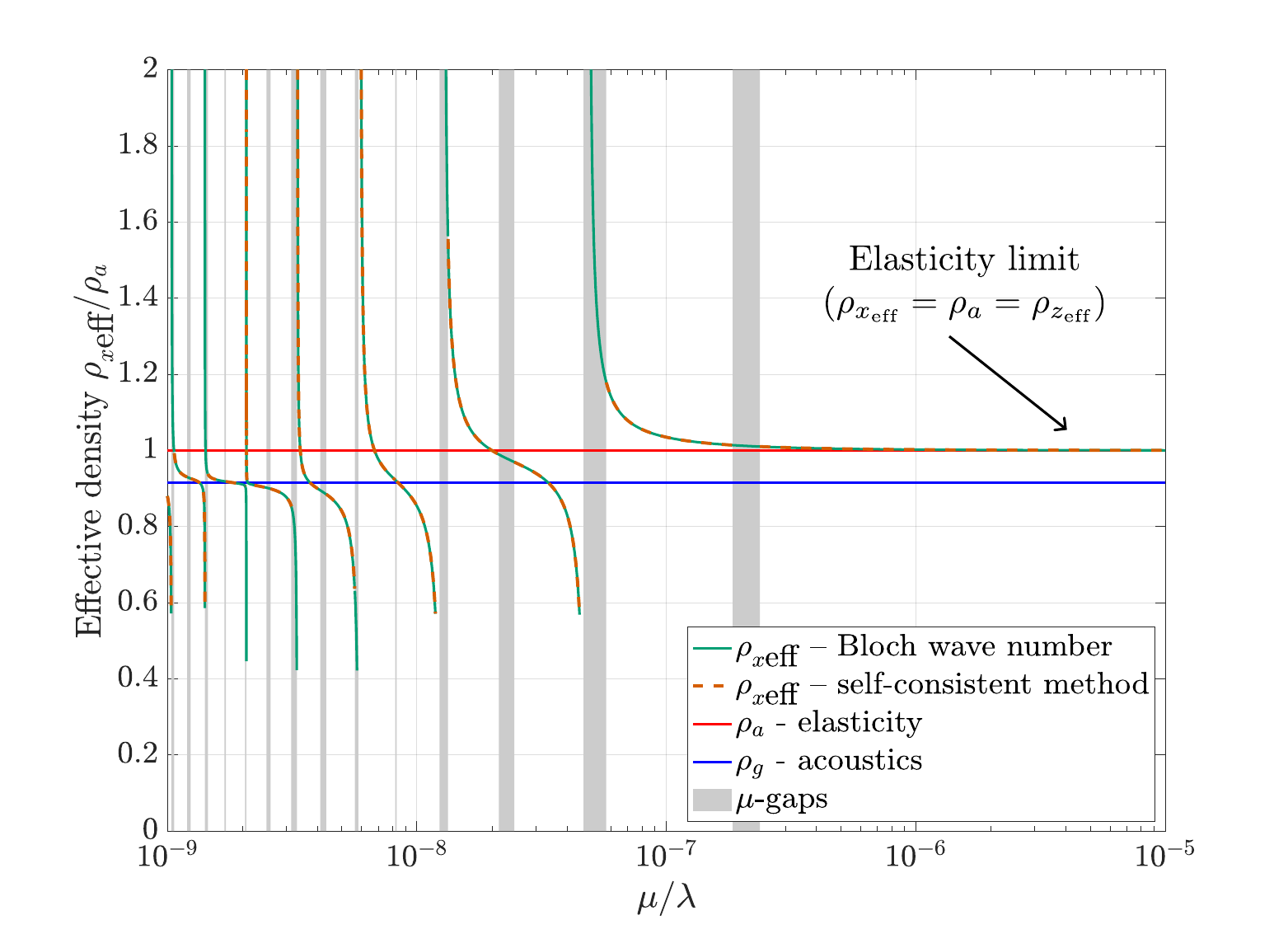}
    \caption{Elasticity to acoustics transition: effective density $\rho_{x \textnormal{eff}}$ as a function of $\mu/\lambda$ obtained via \eqref{Eq_eff_dens_eigs} and calculated via self-consistent method for 5000 unit cells.}
    \label{Fig_elas_eff_dens_bloch_sc}
\end{figure}

Figure \ref{Fig_elas_eff_dens_bloch_sc} shows the effective density $\rho_{x \textnormal{eff}}$ obtained from \eqref{Eq_eff_dens_eigs} compared with $\rho_{x \textnormal{eff}}$ calculated via a dynamic self-consistent method considering a total of 5000 unit cells.
It can be observed that outside of the $\mu$-gaps both effective densities exhibit excellent agreement.
Moreover, the self-consistent method also exhibits the discontinuous behaviour around the BGII-type band gaps observed in Figure \ref{Fig_elas_eff_dens_bloch}. 
This result validates the analytical model in \eqref{Eq_eff_dens_eigs}.
The main discrepancies between the two methods appear in and around the $\mu$-gaps. 
In these regions, \eqref{Eq_eff_dens_eigs} is able to calculate an effective density inside of the $\mu$-gaps while the self-consistent method does not converge.
This can be attributed to the fact that, due to the coupling between compressional and shear modes, the scattering coefficients tend to unity and zero, respectively, therefore preventing the gradient-descent algorithm of being capable of finding the minimum of $\vect{F}$.
Nonetheless, it can be concluded that \eqref{Eq_eff_dens_eigs} is validated for values of $\mu$ located in pass bands where the Bloch wavenumber is real.

\begin{figure}[h!]
    \centering
    \includegraphics[width=\linewidth,trim={0 0.27cm 0 0},clip]{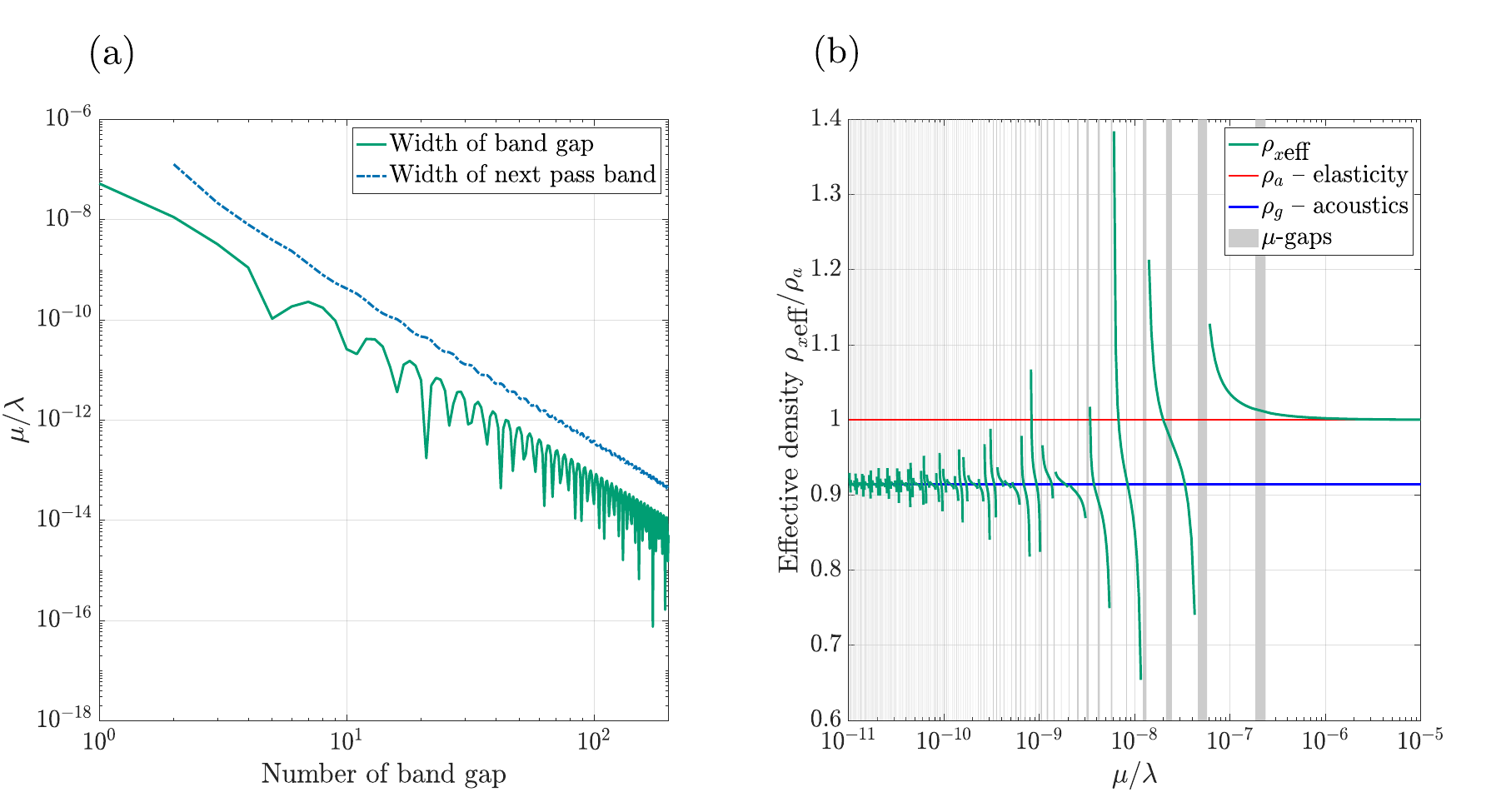}
    \caption{(a) $\mu$-width of band gap (green line) and of the next pass band (blue line) for the first 200 band gaps in the medium. (b) effective density $\rho_{x \textnormal{eff}}$ as a function of $\mu/\lambda$ when $\mu \rightarrow 0$ excluding $\mu$-gaps.}
    \label{Fig_elas_eff_dens_acous_limit}
\end{figure}

So far, the effective density $\rho_{x \textnormal{eff}}$ has been shown for a wide range of values of $\mu$.
However, the examples and discussion presented has not been able to decisively demonstrate the acoustic limit for the effective density.
Thus, the last part of this section deals with said limit.
The main problem that arises when attempting to analytically take $
\mu \rightarrow 0$ for \eqref{Eq_eff_dens_eigs} is the oscillatory behaviour observed in Figures \ref{Fig_elas_eff_dens_bloch} and \ref{Fig_elas_eff_dens_bloch_sc}.
From these figures, as well as Figures \ref{Fig_elas_tm_eigs_mu} and \ref{Fig_elas_tm_bloch_mu}, it is observed that, as $\mu$ decreases, progressively more $\mu$-gaps appear.
This means that, for any very small value of $\mu$, it is always possible to find a smaller value that lies either inside a band gap or in and around EP gaps.
This is problematic, since it was shown that around the BGII-type $\mu$-gaps the effective density exhibits discontinuities where its value tends to diverge.
This leads to the conclusion that it is not possible to take this limit analytically.
Consequently, the following approach is proposed to tackle this problem. 
From the aforementioned figures, it is also straightforward to observe that, as $\mu$ tends to zero, the widths of each $\mu$-gap, as well as those of the adjacent pass band, i.e., the pass band spanning from the left-hand edge of the $j$-th band gap to the right-hand edge of the $j+1$-th band gap, with $j \in \mathbb{N}$, decrease.
Nonetheless, the pass bands appear to be wider than the $\mu$-gaps.
To quantify this, the $\mu$-widths of the first 200 $\mu$-gaps are plotted in Figure \ref{Fig_elas_eff_dens_acous_limit}a.
%
The result displays the width of each $\mu$-gap as well as the width of the subsequent pass band.
It can be clearly seen that, as further $\mu$-gaps are reached, both their widths and the widths of the pass bands decrease, as expected.
Moreover, the plot shows that, for all the band gaps considered, the pass bands are significantly wider than the band gaps.
Therefore, it can be concluded that, as $\mu$ tends to zero, the pass bands dominate over band gaps, their width being, on average, approximately 18 times larger than that of the band gaps.
The previous analysis shows that it is possible to approximate the effective behaviour of the material by the behaviour inside the pass bands as the 'acoustic limit' is taken.
Even so, this result does not give any information about the nature of the dominant behaviour.
To provide a clearer picture of the effective density as $\mu \rightarrow 0$, Figure \ref{Fig_elas_eff_dens_acous_limit}b illustrates the overall behaviour of $\rho_{x \textnormal{eff}}$.
To generate this plot, every pass band is subdivided into a prescribed number of logarithmically-spaced segments. 
The effective density is then evaluated at the boundaries of these segments, excluding the $\mu$-gap edges, as well as at the logarithmic midpoint of each segment. 
The resulting values are plotted against the shear modulus.
In this example, each pass band is divided into three segments, yielding a total of seven evaluation points per pass band.
However, to improve the smoothness of the plot, the first and second pass bands, from right to left in the figure, are subdivided into ten and eight segments, respectively, while the third and fourth are divided into five.
The plot allows to identify both the elastic limit and the acoustic limit directly.
For values of the shear modulus between $\lambda$ and the edge of the first $\mu$-gap, $\rho_{x \textnormal{eff}}$ corresponds to \eqref{Eq_eff_rho_x_elas}, which is the expected result in elasticity.
As further pass bands are investigated, the effective mass density first varies considerably but eventually stabilises around the geometric average, which, as shown previously, corresponds to the acoustic $\rho_{x \textnormal{eff}}$.
Thus, when looking exclusively at the pass band and ignoring the $\mu$-band gaps, the overall behaviour of $\rho_{x \textnormal{eff}}$ tends to its acoustic counterpart as $\mu \rightarrow 0$.

\section{Conclusions}

This work presented the derivation of the effective mass density for the propagation of both acoustic and elastic waves in a infinitely periodic medium.
Particularly, it focused on obtaining an effective description as a function of the shear modulus $\mu$ of the medium, with the goal of considering the limit where $\mu \rightarrow 0$.
For this, acoustic and elastic wave propagation in the unit cell of the medium, comprising two layers of different mass densities and identical elastic properties, was modelled via the transfer matrix method, which allows to straightforwardly consider any number of unit cells.
Moreover, it allowed to analytically calculate the scattering coefficients when placing said medium in between two half-spaces which can exhibit anisotropic mass density.
The transfer matrix approach was then combined with Bloch's theorem to describe wave propagation in a infinitely periodic medium.
This procedure allowed to derive the Bloch wavenumber in the medium from the eigenvalues of the transfer matrix.
Moreover, for elastic waves, it demonstrated the existence of $\mu$-gaps, i.e., ranges of $\mu$ where shear wave propagation in the material is forbidden.
Additionally, due to the interaction between compressional and shear modes brought about by mode conversion in the layer interfaces, exceptional points and avoided crossings were identified around the aforementioned $\mu$-gaps, where the eigenvalues and eigenvectors of both branches coalesce and the modes become highly hybridised.
Lastly, from the Bloch wavenumber it was possible to first derive the effective mass density for acoustics and elasticity, to then find an expression for such effective density that is explicitly dependent on $\mu$, allowing to analyse the limit when the shear modulus tends to zero.

The main result of this work consists in demonstrating the transition from elasticity to acoustics for the effective mass density in a infinitely periodic layered medium, and confirming that the overall behaviour of the isotropic effective mass density in elasticity tends to its anisotropic counterpart in elasticity.
More importantly, the method allowed to derive an analytical expression for such density valid over the entire range of values of $\mu$ where the Bloch wavenumber for compressional waves is real.
Furthermore, it was proven that, even though the effective density for a single unit cell matches the one for the infinitely periodic medium in the limits, the transition between them is heavily modified, particularly by the appearance of the $\mu$-gaps as well as EPs and ACs, which lead to discontinuities in the effective density.
In the EPs, it was demonstrated that, due to mode hybridisation, the wavenumbers for elastic waves become complex, leading to evanescent waves.
This entails that the EP gaps constitute additional physical phenomena, complementary to the traditional band gaps, that lead to forbidden wave propagation in the layered media.
Additionally, it was shown that deriving $\rho_{\text{eff}}$ via the Bloch wavenumber is equivalent to obtaining it via a self-consistent method for a large number of unit cells of a two-layered material placed in between two half-spaces with anisotropic mass density.

In summary, a method for calculating a $\mu$-dependent effective density for acoustic and elastic wave propagation in a layered material through a transfer matrix method combined with Bloch's theorem was introduced in this article.
This approach and the results presented demonstrate the importance of considering said $\mu$-dependence for relating both regimes.
Lastly, a few potential extensions of this work are now presented.
Regarding the existence of the EPs, it would be useful to further understand the physical nature of wave propagation inside the EP gaps, either by numerical or experimental investigation.
The main challenge lies in being able to correctly pinpoint the narrow $\mu$-location of said gaps.
Moreover, an interesting extension could be finding ways of manipulating the EPs, either by, for example, introducing gain or loss in the system, or by considering imperfect boundary conditions.
Additionally, finding a case where an EP coalesces with a frequency band gap edge appears as an interesting example of atypical behaviour for wave propagation in layered media.
Regarding the effective density transition from elasticity to acoustics, a better understanding could be reached by considering 2D- or 3D-periodic materials, such as composites with cylindrical or spherical inclusions, respectively.

\section*{Funding}

G.N. is supported by the Dean’s Doctoral Scholarship Award of The University of Manchester.
\section*{Appendices}
\label{Sec_Append}
\setcounter{subsection}{0}
\renewcommand\thesubsection{\Alph{subsection}}
\subsection{Transfer matrix for an elastic layer} \label{Append_A}

\renewcommand{\theequation}{A.\arabic{equation}}

Each of the terms of the transfer matrix for elastic wave propagation in the two-layered medium are found by computing the transfer matrices for each layer, and then multiplying them to obtain the total transfer matrix.
The terms for a single layer are given by
\begin{alignat}{2}
    T_{11} &= \frac{k_p P_{sx} \xi^p\cos{k_{sz}^*} - k_s P_{px} \xi^s \cos{k_{pz}^*}}{k_p P_{sx} \xi^p - k_s P_{px} \xi^s}, 
    &&T_{12} = \frac{k_p P_{sx} \chi^p \sin{k_{sz}^*} - k_s P_{px} \chi^s \sin{k_{pz}^*}}{i(k_p P_{sz} \chi^p - k_s P_{pz} \chi^s)}, \\
    T_{13} &= i \frac{P_{px} P_{sx} (\cos{k_{sz}^*} - \cos{k_{pz}^*})}{k_p P_{sx} \xi^p - k_s P_{px} \xi^s}, 
    &&T_{14} = \frac{P_{px} P_{sz} \sin{k_{pz}^*} - P_{pz} P_{sx} \sin{k_{sz}^*}}{\mu (k_s P_{pz} \chi^s - k_p P_{sz} \chi^p)}, \\
    T_{21} &= -i \frac{k_p P_{sz} \xi^p \sin{k_{sz}^*} -  k_s P_{pz} \xi^s\sin{k_{pz}^*}}{k_p P_{sx} \xi^p - k_s P_{px} \xi^s}, 
    &&T_{22} = \frac{k_p P_{sz} \chi^p \cos{k_{sz}^*} - k_s P_{pz} \chi^s \cos{k_{pz}^*}}{k_p P_{sz} \chi^p - k_s P_{pz} \chi^s}, \\
    T_{23} &= \frac{P_{px} P_{sz} \sin{k_{sz}^*} - P_{pz} P_{sx} \sin{k_{pz}^*}}{k_p P_{sx} \xi^p - k_s P_{px} \xi^s}, 
    &&T_{24} = i \frac{P_{pz} P_{sz} (\cos{k_{sz}^*} - \cos{k_{pz}^*})}{\mu (k_p P_{sz} \chi^p - k_s P_{pz} \chi^s)}, \\
    T_{31} &= i \frac{k_p k_s \xi^p \xi^s (\cos{k_{sz}^*} - \cos{k_{pz}^*})}{k_p P_{sx} \xi^p - k_s P_{px} \xi^s}, 
    &&T_{32} = \frac{k_p k_s (\xi^s \chi^p \sin{k_{sz}^*} - \xi^p \chi^s \sin{k_{pz}^*})}{k_p P_{sz} \chi^p - k_s P_{pz} \chi^s}, \\
    T_{33} &= \frac{k_s P_{px} \xi^s \cos{k_{sz}^*} - k_p P_{sx} \xi^p \cos{k_{pz}^*}}{k_s P_{px} \xi^s - k_p P_{sx} \xi^p}, 
    &&T_{34} = i \frac{k_s P_{pz} \xi^s \sin{k_{sz}^*} - k_p P_{sz} \xi^p \sin{k_{pz}^*}}{\mu (k_p P_{sz} \chi^p - k_s P_{pz} \chi^s)}, \\
    T_{41} &= \frac{\mu k_p k_s (\xi^p \chi^s \sin{k_{sz}^*} - \xi^s \chi^p \sin{k_{pz}^*})}{k_p P_{sx} \xi^p - k_s P_{px} \xi^s}, 
    &&T_{42} = \frac{i \mu k_p k_s \chi^p \chi^s (\cos{k_{sz}^*} - \cos{k_{pz}^*)}}{k_p P_{sz} \chi^p - k_s P_{pz} \chi^s}, \\
    T_{43} &= \frac{i \mu(k_s P_{px} \chi^s \sin{k_{sz}^*} - k_p P_{sx} \chi^p \sin{k_{pz}^*})}{k_p P_{sx} \xi^p - k_s P_{px} \xi^s}, \quad 
    &&T_{44} = \frac{k_s P_{pz} \chi^s \cos{k_{sz}^*} - k_p P_{sz} \chi^p \cos{k_{pz}^*}}{k_s P_{pz} \chi^s - k_p P_{sz} \chi^p},
\end{alignat}
where $k_{pz}^* = k_p h \cos{\theta_p}$ and $k_{sz}^* = k_s h \cos{\theta_s}$, and the functions $\xi^m$ and $\chi^m$ are given by \eqref{Eq_elas_xi_chi_m} with $m = p$ for compressional and $m =s$ for shear waves.
The angular dependence of each of these functions is omitted to ease notation.
%

\subsection{Derivation of eigenvalues of transfer matrix for elasticity}
\label{Append_B}

\renewcommand{\theequation}{B.\arabic{equation}}

For elasticity, the characteristic equation of the transfer matrix corresponds to a fourth-order polynomial, i.e., four independent eigenvalues, and thus directly solving for its roots becomes challenging.
The problem can be simplified by considering that the transfer matrices of the elastic layers are symplectic \cite{Mead96}, meaning that $\det{(\tensorind{T}{1})} = \det{(\tensorind{T}{2})} = 1$, their characteristic polynomials are reciprocal, and thus their eigenvalues come in two pairs as $\gamma_{1}$ and $\gamma_{1}^\ast = 1/\gamma_{1}$, and $\gamma_{2}$ and $\gamma_{2}^\ast = 1/\gamma_{2}$.
Moreover, this aspect is maintained under product, meaning $\tensor{T} = \tensorind{T}{1} \ \tensorind{T}{2}$ is also symplectic if $\tensorind{T}{1}$ and $\tensorind{T}{2}$ are, and thus $\det{(\tensor{T})} = 1$ and its characteristic polynomial is given by $P(\gamma) = \gamma^4 - c_1 \gamma^3 + c_2 \gamma^2 - c_1 \gamma + 1$, where $c_1$ is the trace of the transfer matrix, and $c_2$ is the sum of the determinants of all the $2x2$ principal sub-matrices of $\tensor{T}$, calculated as in \eqref{Eq_constants_c1_c2}.

The eigenvalues are found by solving $P(\gamma) = 0$.
Due to the symmetry of $P(\gamma)$, the auxiliary variable $\eta = \left( \gamma + 1/\gamma\right)$ is introduced.
Since the eigenvalues of $\tensor{T}$ are either complex conjugate pairs or real reciprocal pairs, $\eta \in \mathbb{R}$.
Dividing $P(\gamma) = 0$ by $\gamma^2$, and considering that $\left( \gamma^2 + 1/\gamma^2\right) = \eta^2 - 2$, it is found that
\begin{equation} \label{Eq_charac_polyn_eta}
    P(\eta) = \eta^2 - c_1 \eta + c_2 - 2 = 0.
\end{equation}
It is straightforward to see that the polynomial has two roots, i.e., $\eta_1$ and $\eta_2$, and subsequently the eigenvalues of the transfer matrix are given by \eqref{Eq_tm_eigs_eta_gamma}.

\subsection{Low-frequency approximation of eigenvalues of acoustic and elastic transfer matrix}
\label{Append_C}

\renewcommand{\theequation}{C.\arabic{equation}}
Considering low-frequency acoustic waves, the eigenvalues of the transfer matrix for acoustic wave propagation in the two-layered medium are given by
\begin{equation} \label{Eq_acous_eig_comp}
    \gamma_1 = 1 - \frac{1}{2} k_{1 z} a k_{2 z} b \left(\frac{r_1}{r_2} + \frac{r_2}{r_1} \right) + \sqrt{\mathcal{S}_a}, \quad
    \gamma_2 = 1 - \frac{1}{2} k_{1 z} a k_{2 z} b \left(\frac{r_1}{r_2} + \frac{r_2}{r_1} \right) - \sqrt{\mathcal{S}_a},
\end{equation}
where $k_{1 z} = k_1 \cos{\theta_1}$ and $k_{2 z} = k_2 \cos{\theta_2}$, and
\begin{equation}
    \mathcal{S}_a = \frac{1}{4} (k_{1 z} a k_{2 z} b)^2 \left(\frac{r_1^2}{r_2^2} + \frac{r_2^2}{r_1^2} - 2 \right) - k_{1 z} a k_{2 z} b \left(\frac{r_1}{r_2} + \frac{r_2}{r_1} \right) - (k_{1 z} a)^2 - (k_{2 z} b)^2 .
\end{equation}

In the case of elastic waves, the eigenvalues of the transfer matrix are derived by considering normal incidence, i.e., $\theta^p_{t_1} = \theta^p_{t_2} = \theta^s_{t_1} = \theta^s_{t_2} = 0$.
Considering low-frequency waves, the eigenvalues are then given by
\begin{align} \label{Eq_eig_m}
    \gamma_{m_1} = 1 - \frac{1}{2}k_{m_1} a k_{m_2}b \left( \frac{k_{m_1}}{k_{m_2}} + \frac{k_{m_2}}{k_{m_1}}  \right) + \sqrt{\mathcal{S}_m}, \quad
    \gamma_{m_2} = 1 - \frac{1}{2}k_{m_1} a k_{m_2}b \left( \frac{k_{m_1}}{k_{m_2}} + \frac{k_{m_2}}{k_{m_1}}  \right) - \sqrt{\mathcal{S}_m}, 
\end{align}
where $m = p,s$ for compressional and shear waves, respectively.
These correspond to two complex-conjugate pairs of eigenvalues, one associated with compressional waves and the other with shear waves.
The term inside the square roots is given by
\begin{equation}
    \mathcal{S}_m =  \frac{1}{4} (k_{m_1} a k_{m_2} b)^2 \left(\frac{k_{m_1}^2}{k_{m_2}^2} + \frac{k_{m_2}^2}{k_{m_1}^2} - 2 \right) - k_{m_1} a k_{m_2} b \left(\frac{k_{m_1}}{k_{m_2}} + \frac{k_{m_2}}{k_{m_1}} \right) - (k_{m_1} a)^2 - (k_{m_2} b)^2 .
\end{equation}

\bibliographystyle{unsrt}
\bibliography{biblio}

\end{document}